\documentclass[fleqn,usenatbib]{mnras}

\usepackage[T1]{fontenc}
\usepackage{ae,aecompl}

\usepackage{graphicx}	% Including figure files
\usepackage{amsmath}	% Advanced maths commands
\usepackage{amssymb}	% Extra maths symbols

\title[GW mergers from wide triple-BHs in the field] {High rate of gravitational waves mergers from flyby perturbations\\ of wide black-hole triples in the field}

\author[E. Michaely and H. B. Perets.]{
Erez Michaely$^{1}$\thanks{E-mail: erezmichaely@gmail.com},
Hagai B. Perets$^{2}$
\\
$^{1}$Astronomy Department, University of Maryland, College Park, MD 20742\\
$^{2}$Physics department, Technion - Israel Institute of Technology, Haifa, Israel 3200002}

\date{Accepted XXX. Received YYY; in original form ZZZ}

\pubyear{2020}

\begin{document}
\label{firstpage}
\pagerange{\pageref{firstpage}--\pageref{lastpage}}
\maketitle

\begin{abstract}
Ultra-wide triple black-holes (TBHs; with an outer orbit $>10^3$ AU) in the field can be considerably
perturbed by flyby encounters with field stars by the excitation of the outer orbit eccentricities. We study the cumulative effect of such flybys,
and show them to be conductive for the production of gravitational-wave (GW) sources. Flyby encounters with
TBHs can turn the TBHs unstable and follow chaotic evolution.
This leads to a binary-single resonant encounter between the outer BH and the inner-binary.
These encounters can result in either a prompt GW-merger of two of the TBH components during the resonant phase, or the disruption of the TBH.
In the latter case a more compact binary is left behind, while the third BH escapes and is ejected.
The compact remnant binary may still inspiral through GW-emission, although on longer timescales.
A significant number of these would lead to a delayed GW-merger in less than a Hubble time.
We find a volumetric merger rate of $\sim3-10{\rm Gpc^{-3}yr^{-1}}$ contributed by the (former) prompt-merger TBH channel 
and $\sim100-250{\rm {\rm Gpc^{-3}yr^{-1}}}$ contributed by the (latter) delayed-merger TBH channel.
The prompt channel gives rise to eccentric mergers in the aLIGO band, while the majority of the delayed-GW
mergers are circularized when enter the aLIGO band. We find the total {\rm eccentric} volumetric
merger rate to be $\sim1-10{\rm Gpc^{-3}yr^{-1}}$ from both channels.
We expect these mergers to show no significant spin-orbit alignment, and uniform delay time distribution.

\end{abstract}
\section{Introduction}
\label{sec:Introduction}
The third observational run (O3) of the aLIGO and VIRGO consortium identified
numerous gravitational-wave (GW) events, the majority of which are
binary black-hole (BBH) mergers. Over the past decades a large body
of theoretical studies was done in order to identify the evolutionary
channels leading to the mergers of two compact objects and predict (before the LIGO era) 
and currently explain the observed rate
of mergers and their properties \citet[e.g.][and more]{Belczynski2002,Belczynski2004,Belczynski2007,Belczynski2008,Belczynski2016,Antonini2012,Dominik2012,Antognini2014,Antonini2014,deMink2015,Petrovich2017}.
In the second observational run (O2) $11$ GW mergers have been detected
by aLIGO and VIRGO. These include $10$ mergers of binary black-holes
(BBHs) and a single merger from a binary neutron-star (NS). All of
the detection were consistent with zero eccentricity %HBP: and with. 
The inferred
BBH-merger rate from O2 (in the local Universe) is $R_{{\rm BBH}}=9.7-101{\rm Gpc^{-3}yr^{-1}}$;
while the merger rates of binary neutron-star is $R_{{\rm BNS}}=110-3840{\rm Gpc^{-3}yr^{-1}}$;
and the upper limit of BH-NS merger is $R_{{\rm BHNS}}<600{\rm Gpc^{-3}yr^{-1}}$
\citep{Abbott2019}.

Four main evolutionary channels were proposed in the context of GW
mergers. The first deals with collisional mergers in dense environments such as
galactic centers or globular clusters \cite[e.g.][]{Rodriguez2016,Rodriguez2018,Fragione2018,Banerjee2018,Hamers2018,Leigh2018,Samsing2014},
where binary mergers are catalyzed by strong interactions with stars
in these dense environment. In such environments, strong binary-single and binary-binary
interactions lead to harden compact binaries (drive them to shorter
periods) and excite their eccentricities. Such models predict GW-production
rates in the range of $2-20{\rm Gpc^{-3}yr^{-1}}.$

The second evolutionary channel deals with the isolated evolution
of initially massive close binary stars \citep[e.g.][]{Belczynski2008,Belczynski2016,Dominik2012,Dominik2015}.
In this scenario massive close binaries strongly interact through one or
two common envelope phases in which the interaction of a star with
the envelope of an evolved companion leads to its inspiral in the
envelope and the production of a short period binary. A fraction of
the post-CE binaries are sufficiently close to merge via GW emission
within a Hubble time. A different merger path is through the \textquotedblleft chemically
homogeneous channel\textquotedblright{} \citep{Mandel2016}. The large
uncertainties in the initial conditions of the binaries, the evolution
in the common-envelope phase, the natal-kick experienced by NS/BHs
at birth; and the mass-loss processes of massive stars give rise to
a wide range of expected GW-sources production rates in the range
$\sim10^{-2}-10^{3}{\rm Gpc^{-3}yr^{-1}}.$

The third evolutionary channel is mergers induced by secular evolution
of triple systems either in the field \citep[e.g.][]{Antonini2016,Antonini2017,Silsbee2017}
or in nuclear cluster and/or massive clusters \citep[e.g.][]{Antonini2012,Petrovich2017,Samsing2018,Hoang2018,Fragione2019,Hamilton2019}.
In this channel the secular and/or semi-secular perturbations by a third companion (Lidov-Kozai
evolution \citet{Lidov1962,koz62}; semi-secular evolution \citet{Antonini2012}) can drive BBHs into high eccentricities
such that they merge within a Hubble-time; the rates expected in this
channel are $\sim0.5-15{\rm Gpc^{-3}yr^{-1}.}$

The fourth channel \citep{Michaely2019} is from wide binaries in
the field (SMA >$1000{\rm AU})$ perturbed by flyby encounters, following similar ideas regarding 
formation of clue stragglers through stellar mergers of wide binaries studied by \citep{Kaib2014}. 
We found that the frequent interactions with random stars can change the eccentricity 
of wide binaries, and in some cases excite sufficiently high eccentricities, 
leading to the merger of the binary via GW emission before the next flyby happens. The predicted
rate from this channel, for spiral galaxies, is $\sim1-10{\rm Gpc^{-3}yr^{-1}.}$ Here we follow up
and extend this channel to study flyby perturbations of wide {\emph triples} in the field. 
%%%
One of the important properties of GW mergers, that can potentially distinguish between the
different channels is the eccentricity of the merged binary in the
aLIGO / VIRGO band. With current observatories only eccentricities
greater than $\sim0.1$ at GW frequency of $\sim10{\rm Hz}$ \citep{Harry2010}
are detectable and termed as ``eccentric mergers''. Currently an
eccentric merger have not been detected yet, and it is important
to understand all evolutionary paths that may lead to an eccentric
merger and their expected rate. It was suggested that eccentric mergers are rare among  mergers of isolated binaries, but that dynamical interaction in dense environments \citep[e.g.][]{Samsing2014,Rodriguez2016} 
could give rise to a non-negligible rate of eccentric mergers.

Another observable property of GW-merges is the measured spin-alignment. Current observations suggest a preference for either an isotropic spin distribution or low spin magnitudes for the observed systems \citep{Will2014}.  Dynamical channels are expected to have isotropic spin orientations, while isolated binaries channels are more likely to have spins that are preferentially aligned with the orbit. As we briefly discuss below, the TBH channel is likely to produce an isotropic distribution (with some possible caveats), more similar to the dynamical channels.

Finally, with the expectations of much more data from the coming runs of aLIGO,  the delay-time
distribution (DTD) of GW mergers could also become an important constraining property. The DTD presents the rate of events since the
formation of BHs or essentially star formation. The mergers time of
BBHs can span from years \citep{Michaely2018} up to Hubble time, with different channels suggesting different DTDs.

In this manuscript we expand our understanding of the fourth channel, and extend it to the study of 
dynamical interactions of wide {\emph triples} in the field (\emph{not} in dense stellar environments).
We calculate the GW merger rate from this channel, characterize the expected eccentricity distribution, and discuss the expected spin-alignment from the channel.

The paper is structured as follows: in section \ref{sec:Wide-triples-in} we
briefly describe the interaction of wide TBH systems in the field
and calculate the rate of these system becoming unstable due to flyby interactions. 
In section \ref{sec:Unstable-triples} we
describe the dynamics of unstable triples and calculate the resulting
galactic GW-merger rate and the rate of eccentric GW mergers. In section \ref{sec:GW-volume-rates}
we compute the corresponding cosmological merger rate observable by LIGO. We discuss our results in 
Section \ref{sec:Discussion} and summarize (section \ref{sec:Summary}).

\section{Wide triples in the field}
\label{sec:Wide-triples-in}
In the following we describe the dynamics of wide triples perturbed by 
random flyby of stars in the field. A more extended mathematical description
of some of the aspects of such interactions an be found in our previous papers \citet{Michaely2016,Michaely2019}.
In what follows we highlight the main aspects of the mathematical
model and key differences of this work focusing on wide triples compared with  \citep{Michaely2019} where we studied wide binaries.
We first describe the interaction qualitatively, subsection \ref{subsec:Qualitative-description}
following with a quantitative treatment in subsection \ref{subsec:Quantitative-description}.

\subsection{Qualitative description}
\label{subsec:Qualitative-description}
\begin{figure}
\includegraphics[width=0.9\columnwidth]{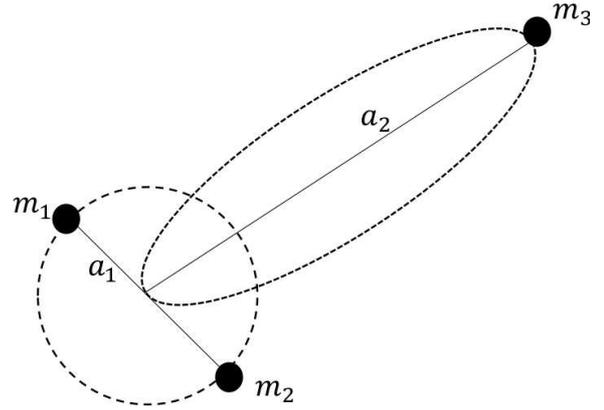}\caption{\label{fig:Illustration-of-hierarchical}Illustration of hierarchical
TBH system, $a_{1}\ll a_{2}$. The inner binary is circular while
the eccentricity of the outer binary is distributed with thermal distribution,
$f\left(e_{2}\right)=2e_{2}.$}

\end{figure}

Several studies (\citet{Kaib2014,Michaely2016,Michaely2019}) showed that the cumulative 
interactions of wide systems
($a\gtrsim1000{\rm AU}$) with field stars through flyby encounters can considerably 
change their (outer-orbit) pericenter distances, mainly through the excitation of the wide-binary eccentricity,
somewhat similar to the case of stars interacting with massive black holes in galactic nuclei  
\citep{Lightman1977,Merritt2013}. A fraction of these
system might interact tidally \citet{Michaely2016} or inspiral through
GW emission \citep{Michaely2019}. Here we focus on wide {\emph triple}-BHs
(TBHs) in hierarchical configurations, where, for simplicity we consider only equal-mass BHs.
In such triples,  the inner binary consists of two components of masses, $m_{1}$ and $m_{2}$,
with the inner orbital parameters; inner semi-major axis (SMA) and inner eccentricity denoted by $a_{1}$ and $e_{1}$, respectively, where, for simplicity we consider only inner binaries with $e_{1}$ set to zero (which might be expected at least for the relatively more compact binaries, if they evolved through a common-envelope evolution phase).
The third BH, $m_{3}$ and the the inner binary (center of mass) serve as 
the outer binary of the triple with the outer SMA denoted by $a_{2}$, where we only consider cases where 
$a_{2}\gg a_{1}$. For illustration see Figure \ref{fig:Illustration-of-hierarchical}.

We note that in this manuscript we are neglecting Lidov-Kozai effects and the effects of mass-loss on such secular evolution
\citep{Lidov1962,koz62,Naoz2016,Michaely2014} because we are focusing
on wide systems with $a_{2}>1000{\rm AU}$. For these systems the
Lidov-Kozai timescale is 
\begin{equation}
\tau_{{\rm LK}}\approx\frac{P_{2}^{2}}{P_{1}}\approx
\end{equation}
 
\[
4.7\cdot10^{12}{\rm yr}\left(\frac{a_{2}}{10^{4}{\rm AU}}\right)^{3}\left(\frac{a_{1}}{0.1{\rm AU}}\right)^{-\frac{3}{2}}\left(\frac{M}{30M_{\odot}}\right)^{-1}\left(\frac{M_{b}}{20M_{\odot}}\right)^{\frac{1}{2}}
\]
where $M\equiv m_{1}+m_{2}+m_{3}$ is the total mass of the TBH and
$M_{b}\equiv m_{1}+m_{2}$ is the total mass of the inner binary, and 
$\tau_{{\rm LK}}$ is typically much larger than a Hubble time, although it could affect TBHs of wider inner binaries and/or closer outer binaries (the latter however, would be less affected by flyby encounters discussed here).
\textbf{In future work we indent to explore the regime where the two timescales overlap and potentially flybys might excite Lidov-Kozai oscillations.}
For these wide systems a flyby can change the eccentricity of the
outer binary such the the pericenter distance, $q=a_{2}\left(1-e_{2}\right)\lesssim a_{1}$.
Namely, the third BH passes within the inner binary SMA, effectively giving rise to a strong binary-single encounter, 
and a chaotic evolution of the now unstable triple. In this case the binary-single encounter resembles the  binary-single 
encounters occurring in dense cluster environments, with similar expected outcomes as those studied in that context (e.g. \citet{Heggie1975,Hills1975,Samsing2014,Stone2019b} and references therein). 
In other words, perturbed wide field triples provide an effective channel of converting isolated field evolution to a 
cluster-like dynamical interaction.

There are two relevant timescales for this part of the model. First,
the interaction timescale, $t_{{\rm int}}\equiv b/v_{{\rm enc}}$,
between the TBH and the flyby field star, where $b$ is the closest
approach of the flyby to the triple system and $v_{{\rm enc}}$ is
the velocity at infinity of the flyby with respect to the triple center
of mass. Second, the outer binary orbital period, $P_{2}$. We restrict
ourselves to the impulsive regime where $t_{{\rm int}}\ll P_{2}$. 
%%%
In the next section \ref{subsec:Quantitative-description} we calculate
the rate of turning hierarchical triples into unstable triples as
a function of the inner SMA.

A fraction of all systems that undergo a dynamical instability phase,
$f_{{\rm merger}}\left(a_{1}\right)$, merge during the resonant interaction
phase; we term this the prompt-merger channel (see section \ref{subsec:Calculating-the-merger}), while the majority are disrupted, with one of the BH ejected, and the other two forming a typically more compact remnant binary. Some of these can later inspiral through GW emission and merge in less than a Hubble time; we term this channel for GW-sources the delayed-merger channel.
For both channels we consider the expected rates, and characterize the expected eccentricity distribution in the LIGO band, and the fraction of eccentric mergers,  $f_{{\rm eccentric}}\text{\ensuremath{\left(a_{1}\right)}. }$ We
find $f_{{\rm merger}}\left(a_{1}\right)$ and $f_{{\rm eccentric}}\text{\ensuremath{\left(a_{1}\right)}  }$
in section \ref{subsec:Calculating-the-merger}.

\subsection{Quantitative description}
\label{subsec:Quantitative-description}
As mentioned earlier, here we briefly review the loss-cone analysis used to estimate the TBH destabilization rates due to flyby encounters. A more detailed discussion of the loss-cone analysis in this context can be found in our previous papers.

Consider a large ensemble of wide TBHs. All BH masses are taken to be equal $m_{1}=m_{2}=m_{3}=10M_{\odot}$
(total mass is denoted by $M$) the inner SMA $a_{1}$ and outer SMA
$a_{2}$. The distribution of $a_{1}$ is log-uniform, $\propto1/a_{1}$
and outer binary SMA $a_{2}>10^{3}{\rm AU}.$ The inner binary is
set to be circular, $e_{1}=0$ while the eccentricity distribution
of outer binary is assumed to be thermal, $f\left(e\right)de=2ede.$. 
The ensemble
is embedded in the field, where the  stellar number density is given by $n_{*}$ and
the typical velocity dispersion, $\sigma_{v}$, is set to be
the relative encounter velocity, $v_{\rm enc}$.

In the following we derive the fraction of the ensemble that sufficiently
interacts with the flyby field stars such that the pericenter of the outer
binary passes within the inner binary SMA, namely $q\leq a_{1}$, potentially a conservative assumption as triple could be destabilized even at larger pericenter separations (e.g. \citet{Mardling2001}). 
We find the fraction dependence on the outer SMA, $a_{2}$ and field
number density, $n_{*}.$ Moreover, we account for outer binary ionization
from the random interaction with flyby stars.

We make use of the loss-cone analysis. We first define the loss cone fraction, $F_{q}$, which is the fraction of systems 
for which $q\leq a_{1}$. The condition of $q=a_{1}$ defines the critical
eccentricity $e_{c}$, where the TBH would destabilize, namely 
\begin{equation}
a_{2}\cdot\left(1-e_{c}\right)=a_{1}
\end{equation}
 which corresponds to $e_{c}=1-a_{1}/a_{2}$. 
\begin{equation}
F_{q}=\int_{e_{c}}^{1}2ede=\frac{2a_{1}}{a_{2}}.
\end{equation}
We note that $F_{q}\ll1$. When a TBH is in the loss cone $m_{3}$
enters the inner binary within a outer orbital period timescale, $P_{2}$
and the triple destabilize it is than lost from the
ensemble of TBHs, and may contribute to the production of GW sources as we discuss below. Systems which orbits are close to the loss cone regime could potentially
be perturbed into it and replenish the loss cone after the next flyby
interaction. In order to calculate what is the fraction of such systems out of
the entire ensemble we calculate the smear cone, the average size
of phase space an outer binary can occupy after an impulsive interaction
with flyby star. The smear cone, defined by $\theta=\left\langle \Delta v\right\rangle /v_{k}$
, where $v_{k}$ is the Keplerian velocity of the outer binary at
the average separation, $\left\langle r\right\rangle =a_{2}\left(1+1/2e^{2}\right)$.
Because $F_{q}\ll1$ we approximate $e\rightarrow1$, namely $v_{k}=\left(GM/3a_{2}\right)^{1/2}$,
where $G$ is Newton's constant. The change in velocity $\Delta v\approx3Ga_{2}m_{p}/v_{{\rm env}}b^{2}$
\citep{Hills1981,Michaely2019} where $m_{p}$ is the mass of the
flyby perturber. Following \citep{Michaely2019}... the size of the
smear cone is 
\begin{equation}
F_{s}=\frac{\pi\theta^{2}}{4\pi}=\frac{27}{4}\left(\frac{m_{p}}{M}\right)^{2}\left(\frac{GM}{a_{2}v_{{\rm enc}}^{2}}\right)\left(\frac{a_{2}}{b}\right)^{4}.\label{eq:SmearCone}
\end{equation}
 The ratio of the smear cone to loss cone is the fraction of the loss
cone filled after a single flyby:
\begin{equation}
\frac{F_{s}}{F_{q}}=\frac{27}{8}\left(\frac{m_{p}}{M}\right)^{2}\left(\frac{GM}{a_{2}v_{{\rm enc}}^{2}}\right)\left(\frac{a_{2}}{b}\right)^{4}\left(\frac{a_{2}}{a_{1}}\right).
\end{equation}
In the case where the loss cone is continuously fully replenished, 
$F_{q}=F_{s}$,
the timescale for the loss-cone replenishment becomes comparable to the timescale for the loss-cone depletion, i.e. the outer orbit orbital period, $P_{2}$. Therefore the rate of depletion which is
a function of the size of the loss cone is
\begin{equation}
\dot{L}_{{\rm Full}}=\frac{F_{q}}{P_{2}}\propto a_{2}^{-5/2}a_{1},
\end{equation}
which is independent of the local stellar density, $n_{*}$ and scales
linearly with the inner binary SMA, $a_{1}$. Therefore the depletion
rate decreases with increasing outer SMA in the full loss cone regime.

On the other hand, for the case where $F_{s}<F_{q}$, namely for tighter outer
binaries, which are less susceptible for change due to random flyby
interaction (\ref{eq:SmearCone}), the loss cone is not completely
full at all times, and one needs to consider the so called empty loss cone regime. In this case
the depletion rate depends on the rate of systems being kicked into
the loss cone. Specifically, $f=n_{*}\sigma v_{{\rm enc}}$ where
$\sigma=\pi b^{2}$ is the geometric cross-section of the random flyby
interaction. In this case the typical timescale for  the depletion is the timescale for entering the loss cone, namely $T_{{\rm empty}}=1/f$.
As we showed previously, $f$ can be written as 
\citep{Michaely2016,Michaely2019} 
\begin{equation}
f=n_{*}\pi\sqrt{\frac{27}{8}\left(\frac{m_{p}}{M}\right)^{2}\frac{GMa_{2}^{4}}{a_{1}}}.
\end{equation}
The critical SMA for which the two timescales are equal the depletion rate is equal to the rate of systems entering the loss cone \citep{Michaely2019} is given by
\begin{equation}
a_{{\rm crit}}=\left(\frac{2}{27\pi^{4}}\frac{M}{m_{p}^{2}}\frac{a_{1}}{n_{*}^{2}}\right)^{1/7}.
\end{equation}
 Using $a_{{\rm crit}}$ we can calculate the fraction of systems that
enter the loss cone for both regimes: $a<a_{{\rm crit}}$the empty
loss cone; $a>a_{{\rm crit}}$ the full loss cone.

\begin{figure}
\includegraphics[width=0.8\columnwidth]{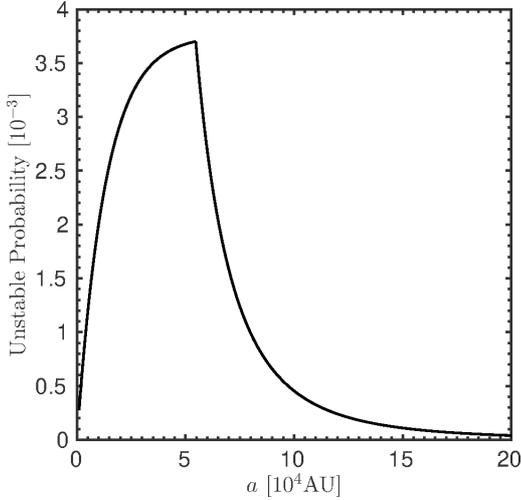}\caption{\label{fig:Probability-of-becoming}The probability of becoming unstable,
namely the outer pericenter distance $q_{2}=a_{2}\left(1-e_{2}\right)\protect\leq a_{1}$,
due to flyby interaction. The plot is calculated for the following
parameters: $t=10{\rm Gyr}$, $n_{*}=0.1{\rm pc^{-3}},$ $v_{{\rm enc}}=50{\rm kms^{-1}}$.
The highest probability is for $a_{{\rm crit}}$. The full loss cone
regime is for $a>a_{{\rm crit}}$ and the empty loss cone regime is
for $a<a_{{\rm crit}}.$}

\end{figure}

The loss cone, $F_{q}$ represents the fraction of systems that are
lost from the ensemble after the relevant timescale, therefore $\left(1-F_{q}\right)$
is the surviving fraction. For the empty loss cone regime this timescale
is $T_{{\rm empty}}=1/f$ while for the full loss cone the timescale
is $P_{2}$. We can write the fraction of systems that enter the loss
cone as function of time, $t$ as 
\begin{equation}
L\left(a_{1},a_{2},n_{*}\right)_{{\rm empty}}=1-\left(1-F_{q}\left(a_{1},a_{2}\right)\right)^{t\cdot f}.
\end{equation}
At the limit where $F_{q}t/T_{{\rm empty}}\ll1$ we can expand this equation
to the leading order and get 
\begin{equation}
L\left(a_{1},a_{2},n_{*}\right)_{{\rm empty}}=F_{q}tf.
\end{equation}
Note that the fraction of systems lost in the empty loss cone regime
is proportional to $F_{q}$, namely 
\begin{equation}
L_{{\rm empty}}\propto F_{q}\propto a_{2}^{-1}a_{1}.
\end{equation}
Specifically, the fraction grows with SMA for $a_{2}<a_{{\rm crit}}$,
unlike the full loss cone regime. This means that the loss-rate peaks 
for TBHs with SMA of $a_{{\rm crit}}.$ For the full
loss cone we follow the same treatment with 
\begin{equation}
L\left(a_{1},a_{2},n_{*}\right)_{{\rm full}}=1-\left(1-F_{q}\left(a_{1},a_{2}\right)\right)^{t/P_{2}},
\end{equation}
and after the expansion we get 
\begin{equation}
L\left(a_{1},a_{2},n_{*}\right)_{{\rm full}}=F_{q}t/P_{2}.
\end{equation}
Our treatment so far neglected the ionization process for wide systems
in collisional environments. Taking the ionization into account by
using the half-life time treatment from \citep{Bahcall1985}, where
the half life time is defined to be 
\begin{equation}
t_{1/2}=0.00233\frac{v_{{\rm enc}}}{Gm_{p}n_{*}a_{2}}
\end{equation}
we get for the empty loss cone
\begin{equation}
L\left(a_{1},a_{2},n_{*}\right)_{{\rm empty}}=\tau F_{q}f\left(1-e^{-t/\tau}\right)=\label{eq:empty}
\end{equation}
\[
\tau\frac{2a_{1}}{a_{2}}n_{*}\pi\sqrt{\frac{27}{8}\left(\frac{m_{p}}{M}\right)^{2}\frac{GMa_{2}^{4}}{a_{1}}}\left(1-e^{-t/\tau}\right).
\]
where $\tau=t_{1/2}/\ln2$. For the full loss cone we get 
\begin{equation}
L\left(a_{1},a_{2},n_{*}\right)_{{\rm full}}=\tau\frac{F_{q}}{P_{2}}\left(1-e^{-t/\tau}\right)=\label{eq:full}
\end{equation}
\begin{equation}
\tau\frac{2a_{1}}{a_{2}}\left(\frac{GM}{4\pi^{2}a_{2}^{3}}\right)^{1/2}\left(1-e^{-t/\tau}\right).
\end{equation}
We emphasize the fact that in both regimes the loss-cone fraction is proportional
to the inner SMA $a_{1}$. We can identify the loss-fraction to be
the probability for a TBH to become unstable due to flyby interactions.
Figure \ref{fig:Probability-of-becoming} shows a representative case
for the probability of becoming unstable as a function of the outer SMA
for some specific time during the evolution and for a specific given field environment.

Equipped with these equations we turn to calculate the fraction of GW-mergers that occur following the strong encounter between the outer third companion and the inner binary (i.e. the now unstable triple) catalyzed by the flyby perturbations. 

\section{Unstable triples}
\label{sec:Unstable-triples}
In this section we describe the dynamics of unstable tripes. We follow
closely the treatment done by \citep{Samsing2014,Samsing2018b} in the context of binary-single encounters.
It is well known that triple systems are not believed to be integrable
and therefore we cannot predict the end result of any specif triple
system. However, in a statistical manner we can predict the end state
of binary-single encounter \citep{Stone2019b,Samsing2014,Heggie1975}.

Binary-single encounters are an important astrophysical source of
unstable triples. The physics of binary-single encounters were studied
mainly in dense stellar environment such as globular clusters or galactic
nuclei. A close binary-single interaction is considered when the
single star passes within the binary SMA, or specifically within the sphere
of influence of the binary. In this situation the gravitational interaction
between every pair of masses is comparable in strength and the outcome
is chaotic. For such close interactions two outcomes are possible. The
first, direct interaction (DI) where only one gravitational interaction
takes place and the result is a tighter binary and an escaper. Note
the binary could be either the same as in the initial condition, this
case is call a \textit{flyby}, or different and this case is called
an \textit{exchange}. The second, intermediate state (IMS), where
the systems goes through many (of the order of $\left\langle N_{{\rm IMS}}\right\rangle =20$, 
for our case \citep{Samsing2017}) binary-single encounters, where each time
the orbital characteristics (SMA and eccentricity) are drawn from
the available phase space volume set by the system angular momentum and
energy budget. Keep in mind that when the binary orbital properties are
set, conservation of angular momentum and energy set the trajectory
of the bound third star until the next binary-single scatter. The
end-state of the multiple binary-single scattering is a tight binary
and an escaper.

From the GW perspective a merger can occur either {\rm promptly} between scattering events during the IMS, or later, after an end-state is reached, when one of the BHs escapes, leaving behind a more compact, likely eccentric binary.
The remnant binary would eventually inspiral and merge through GW-emission on a typically much longer timescale than the dynamical time.  A fraction of latter mergers occur in less than a Hubble time and these would contribute to the rate of detectable GW sources; we term this GW-sources channel the {\rm delayed-merger} channel. In the following
we calculate the rate of mergers and eccentricity distribution of
the merged systems in both cases.

\subsection{Binary-single encounters and the production of prompt GW-mergers}

In this subsection we describe the mathematical modeling of the IMS.
In the following we consider only equal masses BH with $10M_{\odot}$each.
The initial binary is circular with SMA, $a_{1}$, and the third BH interacts
with the binary via consecutive binary-single encounters. In each
encounter the probability for forming a temporary binary with any two
out of the three BHs is uniform. The eccentricity, $e_{{\rm IMS}}$
is drawn from thermal distribution, namely $f\left(e\right)de=2ede$.
The SMA is determined by the energy budget which is approximated by
equation 12 in \citep{Samsing2018b} 
\begin{equation}
\frac{m_{1}m_{2}}{2a_{1}}=\frac{m_{i}m_{j}}{2a_{{\rm IMS}}}+\frac{m_{ij}m_{k}}{2a_{{\rm bs}}}\label{eq:energy budget}
\end{equation}
where $a_{{\rm IMS}}$ is the SMA of the temporary binary and $a_{{\rm bs}}$
is the temporary SMA of the outer binary. Where $\left\{ i,j,k\right\} $
are the randomized indexes after the interaction and $m_{ij}=m_{i}+m_{j}$
is the mass of the temporary binary. From eq. (\ref{eq:energy budget})
we can express the SMA of the third bound BH
\begin{equation}
a_{{\rm bs}}=a_{1}\left(\frac{m_{ij}m_{k}}{m_{1}m_{2}}\right)\left(\frac{a'}{a'-1}\right)\label{eq:a_bs}
\end{equation}
where 
\begin{equation}
a'\equiv\frac{a_{{\rm IMS}}}{a_{c}}\ {\rm and}\ a_{c}\equiv a_{1}\frac{m_{i}m_{j}}{m_{1}m_{2}}.\label{eq:a_definitions}
\end{equation}
We note that in our equal mass case $a_{c}=a_{1}$ and therefore $a'$
is just $a_{{\rm IMS}}/a_{1}$.

In order to estimate the available phase space for the IMS we estimate
the upper (lower) bound $a'_{{\rm U}}\ \text{\ensuremath{\left(a'_{{\rm L}}\right)}}$
of $a'$. The lower bound of $a'$ is trivial with 
\begin{equation}
a_{{\rm L}}'\approx1,\label{eq:a_L}
\end{equation}
the upper bound should separate between when the resonant triple can
no longer be described as an IMS (a binary and a bound single), this
occurs when $a_{{\rm bs}}\approx a_{{\rm IMS}}$. \citet{Samsing2018b}
finds that one way of estimating $a'_{{\rm U}}$ is by comparing the
tidal force, $F_{{\rm tid}}$ exerted by the third BH with the binary
gravitational binding force, $F_{{\rm bin}}$. In the high eccentricity
limit we find 
\begin{equation}
F_{{\rm tid}}\approx\frac{1}{2}\frac{Gm_{ij}m_{k}}{a_{{\rm bs}}^{2}}\frac{a_{{\rm IMS}}}{a_{{\rm bs}}}
\end{equation}
\begin{equation}
F_{{\rm bin}}\approx\frac{1}{4}\frac{Gm_{i}m_{j}}{a_{{\rm IMS}}^{2}}.
\end{equation}
We set $a'_{{\rm U}}$ in the case that 
\begin{equation}
\frac{F_{{\rm tid}}}{F_{{\rm bin}}}=0.5
\end{equation}
which translates to 
\begin{equation}
a'_{{\rm U}}=1+\left(\frac{1}{2}\frac{m_{k}}{\mu_{{\rm ij}}}\right)^{2/3}\label{eq:a_U}
\end{equation}
where $\mu_{ij}\equiv m_{i}m_{j}/(m_{i}+m_{j})$ is the reduced mass
of the IMS binary.

The values of $a'$ are distributed uniformly between $a'_{{\rm L}}$
and $a'_{{\rm U}}$ and the eccentricity distribution is thermal \citep{Heggie1975,Hut1985,Rodriguez2018}.

Next we can calculate the orbital timescale for the third companion, $t_{{\rm iso}}$, to come back for the next binary-single encounter. During the tie in-between scatter events the
temporary binary can potentially merge via GW emission if its merger timescale,
$t_{{\rm merger}}$ is shorter than $t_{{\rm iso}}$. 

The orbital
period is simply the Keplerian orbital period with $a_{{\rm bs}}$;
combining it with eq. (\ref{eq:a_bs}) and eq. (\ref{eq:a_definitions})
we get:
\begin{equation}
t_{{\rm iso}}=2\pi\frac{a_{1}^{3/2}}{\sqrt{GM}}\left(\frac{m_{ij}m_{k}}{m_{1}m_{2}}\right)^{3/2}\left(\frac{a'}{a'-1}\right)^{3/2}.\label{eq:t_iso}
\end{equation}
The merger timescale, for eccentric binaries, is given by \citep{Pet64}
\begin{equation}
t_{{\rm merger}}\approx\frac{768}{425}T_{c}\text{\ensuremath{\left(a_{{\rm IMS}}\right)}}\left(1-e_{{\rm IMS}}^{2}\right)^{7/2}\label{eq:t_merger}
\end{equation}
where $T_{c}=a_{{\rm IMS}}^{4}/\beta$ is the merger timescale for
a circular orbit and $\beta=64G^{3}m_{i}m_{j}\left(m_{i}+m_{j}\right)/\left(5c^{2}\right)$
where $c$ is the speed of light.

\subsubsection{Calculating the merger fraction}
\label{subsec:Calculating-the-merger}
\begin{figure}
\includegraphics[width=0.9\columnwidth]{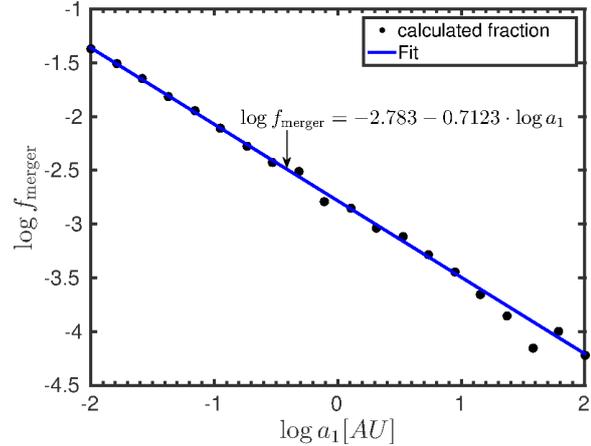}\caption{\label{fig:f_merger} The fraction of systems that merged, $f_{{\rm merger}}$during
the resonant phase as a function of the initial SMA, $a_{1}$. For
every $a_{1}$ we simulated $10^{5}$ binary-single scattering experiments;
for each experiment we use $N_{{\rm IMS}}=20$ scattering events which
we randomize the temporary binary orbital elements (see text) and
check if this temporary IMS leads to a merger. Black dots, the calculated
fraction from our numerical experiment. Blue solid line, the best
fit to a power law.}
\end{figure}

In order to find the fraction of systems that merge during the IMS
as a function of the initial SMA, $a_{1}$, \textbf{we preform a
numerical calculation. In order to save computer time we do not make a direct N-body
simulation. We sample, in a Monte-Carlo approach, the IMSs orbital distributions from \citep{Samsing2014,Samsing2017} and check whether or not they lead to a merger. We use MATLAB to} sample 20 values of $a_{1}$ equally spaced
in log from $\left(10^{-2}{\rm AU},10^{2}{\rm AU}\right)$. For each
value of $a_{1}$ we simulate $N_{{\rm tot}}=10^{5}$ scattering experiments
where for each scattering experiment there is $N_{{\rm IMS}}=20$
times where a temporary binary is created bound to a third BH on a
Keplerian orbit. For each iteration of the IMS we randomly choose the binary orbital
properties, $a_{{\rm IMS}}$, which are  drawn from a uniform distribution in the range $\left(a'_{{\rm L}},a'_{{\rm U}}\right)$
see equations (\ref{eq:a_L}) and (\ref{eq:a_U}); and the eccentricity,
$e_{{\rm ecc}}$ is drawn from a thermal distribution. Next, we calculate
$t_{{\rm iso}}$ from eq. (\ref{eq:t_iso}) and compare it to $t_{{\rm merger}}$
from eq. (\ref{eq:t_merger}). If $t_{{\rm merger}}<t_{{\rm iso}}$
we count it as an IMS merger and check whether it is an eccentric
merger in the LIGO band, see subsection \ref{subsec:Calculating-eccentric-mergers}.
If $t_{{\rm merger}}>t_{{\rm iso}}$ we randomize the binary and single
again until we reach $N_{{\rm IMS}}$ times. Additionally we record
all $t_{{\rm iso}}$ in order to calculate the merger time since the
beginning of the scattering experiment. In the case where no merger occurs during
the resonant phase we record the final end state, to eventually obtain the distribution of the orbital parameters from such cases (see subsection \ref{subsec:Post-resonance-state}).
$f_{{\rm merger}}\left(a_{1}\right)$ is then just the number of mergers
divided by the total number of systems considered, $N_{{\rm tot}}$. 
The results presented in Figure \ref{fig:f_merger}.
We find a power law relation between $f_{{\rm merger}}$ and $a_{1}$,
the exact fitted function is 
\begin{equation}
f_{{\rm merger}}\left(a_{1}\right)=0.00165\times a_{1}^{-0.7123}.\label{eq:merger_fit}
\end{equation}
We note that the fraction scales with the inner SMA with a power which
is smaller than unity $f_{{\rm merger}}\propto a_{1}^{-0.7123}.$

\subsubsection{Calculating the fraction of eccentric mergers }
\label{subsec:Calculating-eccentric-mergers}
In order to find $f_{{\rm eccentric}}\text{\ensuremath{\left(a_{1}\right)} }$
we simulate the evolution of each binary we flagged as an IMS merger.
We use the well known equations of motion of the SMA and eccentricity
from \citep{Pet64} 
\begin{equation}
\frac{da}{dt}=-\frac{64}{5}\frac{G^{3}m_{1}m_{2}\left(m_{1}+m_{2}\right)}{c^{5}a^{3}\left(1-e^{2}\right)^{7/2}}\left(1+\frac{73}{24}e^{2}+\frac{37}{96}e^{4}\right)
\end{equation}
\begin{equation}
\frac{de}{dt}=-e\frac{304}{15}\frac{G^{3}m_{1}m_{2}\left(m_{1}+m_{2}\right)}{c^{5}a^{4}\left(1-e^{2}\right)^{5/2}}\left(1+\frac{121}{304}e^{2}\right).
\end{equation}

\begin{figure}
\includegraphics[width=0.9\columnwidth]{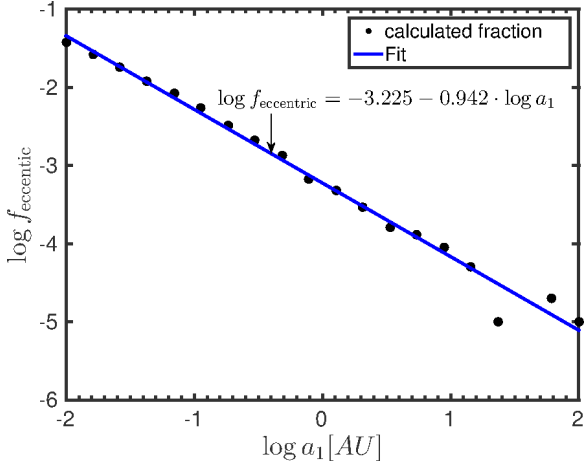}\caption{\label{fig:Fraction_eccentric}The fraction of eccentric mergers,
$f_{{\rm eccentric}}$ during the resonant phase as a function of
the initial SMA, $a_{1}$. For the same setup as in Figure \ref{fig:f_merger}.
Black dots, the calculated fraction from our numerical experiment.
Blue solid line, the best fit to a power law.}
\end{figure}

\begin{figure}
\includegraphics[width=1.0\columnwidth]{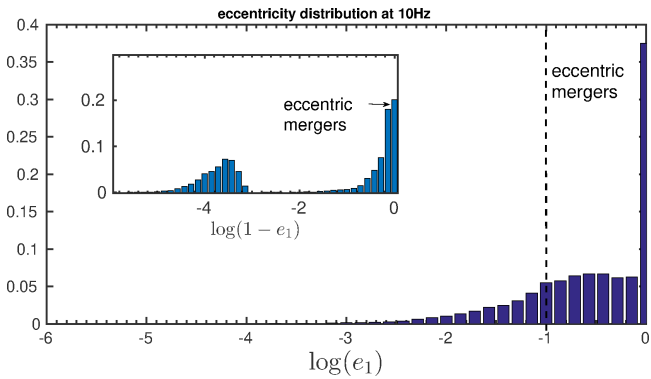}\caption{\label{fig:ecc_distribution}The eccentricity distribution at $10{\rm Hz}$
GW frequency of our entire sample. The main plot show the distribution
of $\log e$ , hence all values greater that $-1$ are eccentric mergers,
$\sim78\%$. This distribution is weighed with the distribution
of the initial SMA, $a_{1}.$ The inset is the same distribution but
presented in $\log\left(1-e_{1}\right)$, namely focuses on the most
eccentric part of the distribution we see the most probable $e$ value
correspond to $\log\left(1-e_{1}\right)\approx-3.5$.}
\end{figure}

Additionally we calculate the approximate gravitational peak frequency
following \citep{Wen2003} 
\begin{equation}
f_{{\rm peak}}\text{\ensuremath{\left(a_{{\rm IMS}},e_{{\rm IMS}}\right)}}=\frac{1}{\pi}\sqrt{\frac{Gm_{ij}}{a_{{\rm IMS}}^{3}}}\frac{\left(1+e_{{\rm IMS}} \right)^{1.1954}}{\left(1-e_{{\rm IMS}}^{2}\right)^{1.5}}.
\end{equation}
We consider a binary merger to be an eccentric merger if the eccentricity, $e_{{\rm IMS}}$
is greater than $0.1$ when the GW peak frequency is $f_{{\rm peak}}=10{\rm Hz}$
\citep[e.g.][]{Rodriguez2018}. In Figure \ref{fig:Fraction_eccentric}
we present our calculated eccentric merger fraction as a function
of the initial SMA, $a_{1}$. The relation between the eccentric fraction
and the initial SMA is well described by a power law 
\begin{equation}
f_{{\rm eccentric}}\left(a_{1}\right)=0.0006\times a_{1}^{-0.942}.\label{eq:eccentric_fit}
\end{equation}
In Figure \ref{fig:ecc_distribution} we present the eccentricity
distribution at $10{\rm Hz}$ for the entire sample weighted by the inner SMA distribution.
We report that $\sim78\%$ of all mergers in the IMSs are eccentric
in the aLIGO band.

\subsection{The post encounter states and the production of delayed-mergers}
\label{subsec:Post-resonance-state}
In the following we study the the production of delayed mergers in cases where no prompt merger occurs during the resonant encounter, and a remnant compact binary is formed with $a_{{\rm delay}}<a_{1}$ (while the third BH is ejected from the system). It was shown in \citep{Stone2019b}
that the energy distribution of the remnant binary scales like 
\begin{equation}
E_{{\rm delay}}\propto\left|E_{1}\right|^{-4}
\end{equation}
where $E_{{\rm delay}}$ is the energy of the remnant binary and $E_{1}=-Gm_{1}m_{2}/(2a_{1})$
is the initial binary energy. 

Additionally, the eccentricity of the remnant binary,
$e_{{\rm delay}}$, is drawn from thermal distribution \citep{Stone2019b}.

\begin{figure*}
\includegraphics[width=1.04\columnwidth]{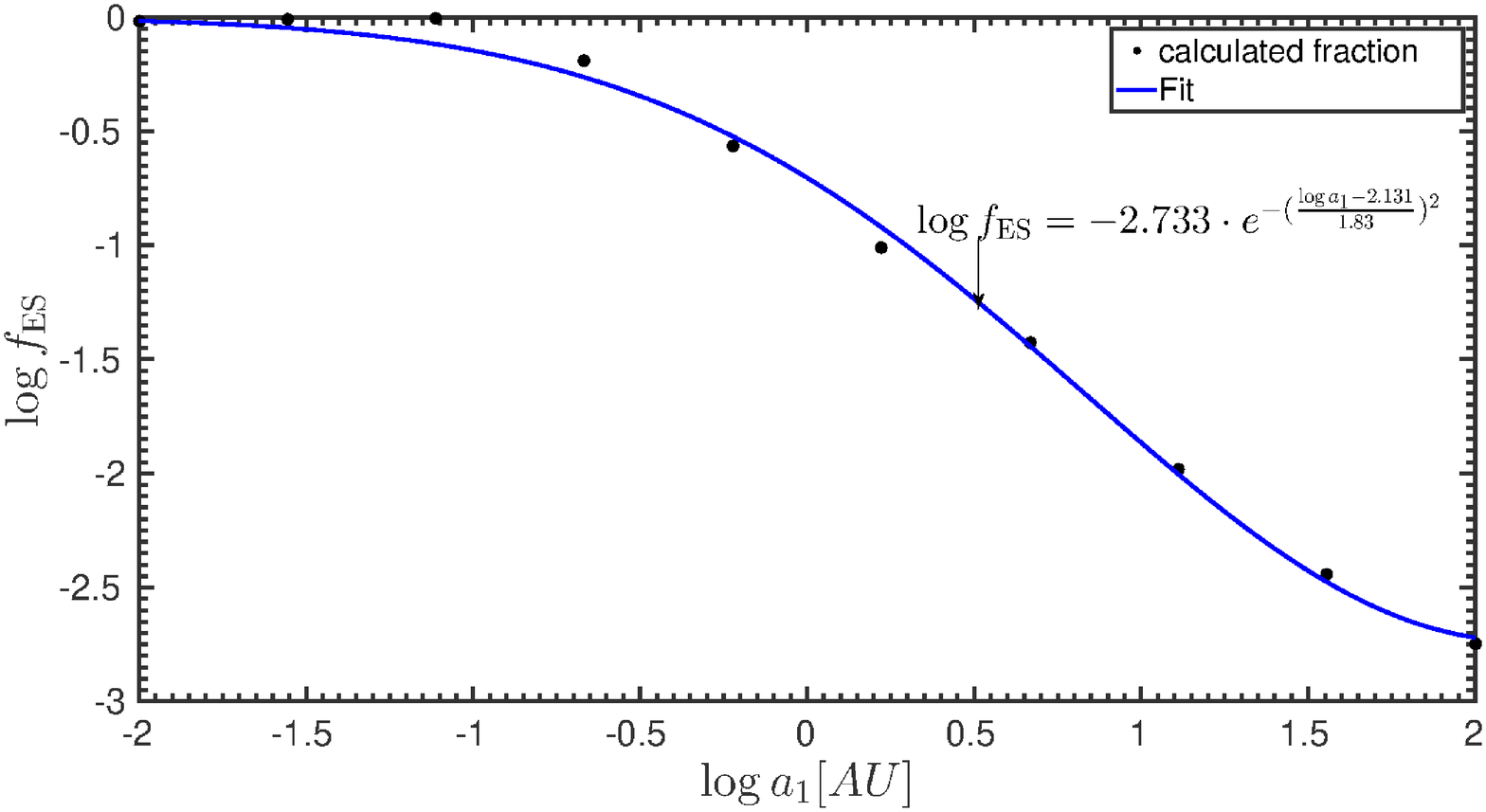} \includegraphics[width=1.04\columnwidth]{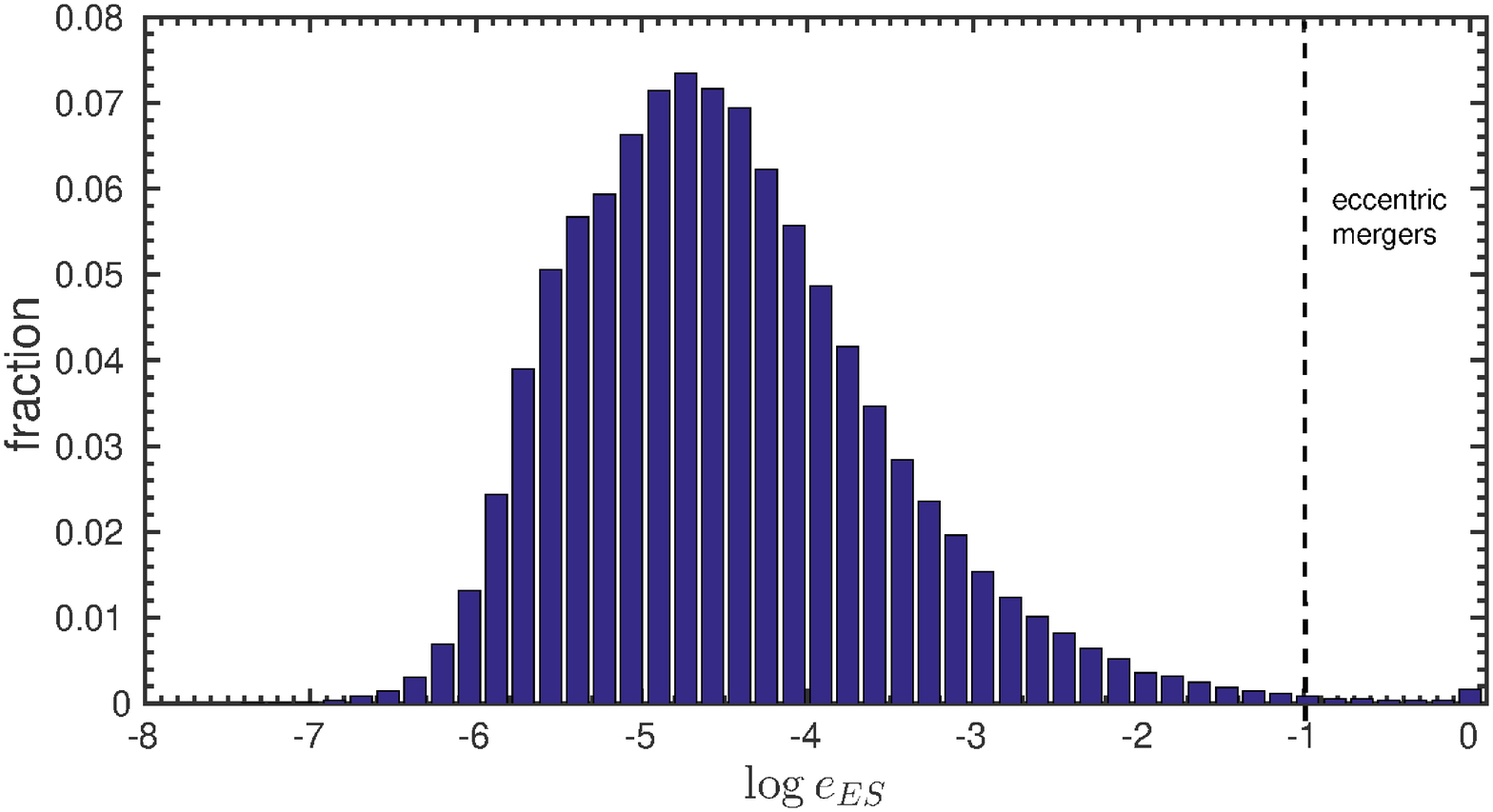}\caption{\label{fig:ecc_dist_endstate}Left plot: The fraction of delayed
mergers, $f_{{\rm delay}}$ as a function of the initial SMA, $a_{1}$
out of $10^{4}$ binary-single experiments. A merger is flagged when
the merger timescale is less than the Hubble time, $t_{{\rm Hubble}}.$
Black dots, the calculated fraction from our numerical experiment.
Blue solid line, the best fit to a Gaussian with the variable, $\log a_{1}$.
Right plot: The eccentricity distribution at $10{\rm Hz}$ for all
delay mergers, weighted with the initial SMA, $a_{1}$distribution. The
plot is similar to \citep{Rodriguez2016,Samsing2017}. Only $\sim0.3\%$
of the entire population is eccentric at $\sim10{\rm Hz}$}

\end{figure*}

For every system that did not promptly merge during the IMS we
follow the GW-inspiral evolution of the remnant binary and calculate the
merger time by using eq. (\ref{eq:t_merger}), and the eccentricity at $f_{{\rm peak}}=10{\rm Hz}$ by following the same treatment as
described in (\ref{subsec:Calculating-eccentric-mergers}) for the prompt mergers, where the only difference is such binaries are followed up to a Hubble time, as their evolution is not restricted by the next encounter (i.e. at the isolation time). The left panel of Figure
\ref{fig:ecc_dist_endstate} shows the fraction of systems
that merge within a Hubble time, and the eccentricity distribution for our sample,  weighted by the distribution of the inner SMA, is presented
in the right panel. We note that the eccentricity distribution is
very similar to \citep{Rodriguez2018,Samsing2018c}, besides the somewhat lower fraction of eccentric mergers, as these mergers do not include the prompt mergers discussed before. The combined distribution of both the prompt and delayed mergers is comparable with those found by \citep{Rodriguez2018}, as might be expected given that the final distribution in both cased is generally determined by the outcomes of binary-single encounters.
    
We found a numerical
Gaussian fit to the merger fraction of the delayed mergers 
\begin{equation}
\log f_{{\rm delay}}=-3.237\times e^{-\left(\frac{\log a_{1}-2.212}{1.964}\right)^{2}}, \label{eq:f_delay}
\end{equation}
 and find that the total fraction of eccentric delayed mergers is $f_{{\rm delay,ecc}}\approx1\%$.

We note that the majority of the prompt mergers are eccentric
because the merger time is limited to the isolation time $t_{{\rm iso}}\ll t_{{\rm Hubble}}$, and only the most initially eccentric binaries, having small peri-centers, could merge on such short timescales. In contrast, we find a smaller fraction of eccentric delayed mergers  because the merger time
is instead limited by $t_{{\rm Hubble}}$, allowing for binaries that merge 
on these longer timescales to also circularize by the time they reach the 
aLIGO band. 

We emphasize that isolated binaries with SMA smaller than $0.1{\rm AU}$
will merge withing a Hubble time, therefore we expect the fitted function
from equation (\ref{eq:f_delay}) to saturate at the lower end of $a_{1}$
near unity. The merger rate we calculate for these binaries
is therefore effectively included in the isolated binary channels discussed by others \citep[and others]{Belczynski2016,Belczynski2008,Dominik2012}. Nevertheless, he IMS evolution could give rise to higher initial eccentricity of such binaries and thereby (on average) shorter merger time and potentially higher detected eccenttricities, even if not affecting the total merger rate. Although this is generally the case, in terms of longer delay times, the relatively low fraction of eccentric mergers coming from these (small initial SMA) mergers indicates that these mergers are effectively indistinguishable from the isolated binary case at least in the aLIGO band. One should note that they may still present significantly higher eccentricities at earlier stages, potentially observable by future space-based GW-detectors. The latter aspects, however, are beyond the scope of the current paper.

\section{Volumetric rates of GW-sources from the TBH channel}
\label{sec:GW-volume-rates}
In this section we calculate the volumetric rate of GW mergers from TBHs in both the prompt and delayed mergers channels. 
We make use of the equations (\ref{eq:empty})
and (\ref{eq:full}) to calculate the fraction of TBHs that become
unstable due to flybys, and then combine them with the merger fractions described above, $f_{{\rm merger}}$ and
$f_{{\rm ecc}}$ (from equations (\ref{eq:merger_fit}) and (\ref{eq:eccentric_fit})).
Since the different properties of spiral and elliptical galaxies affect the rates (through the different stellar number density and velocity dispersions in the different types of galaxies), we calculate the merger rate for both typical spiral and elliptical galaxies. For the
spiral galaxy case we model a Milky Way (MW)--like galaxy stellar
density similar to that considered in \citep{Michaely2016}. Let $dN(r)=n_{*}\left(r\right)\cdot2\pi\cdot r\cdot h\cdot dr$
be the the number of stars in a region $dr$ (and scale height $h$),
located at distance $r$ from the center of the Galaxy. Additionally,
we model the Galactic stellar density in the Galactic disk as follows
\begin{equation}
n_{*}\left(r\right)=n_{0}e^{-\left(r-r_{\odot}\right)/R_{l}}\label{eq:MW_galaxy}
\end{equation}
, where $n_{0}=0.1{\rm pc}^{-3}$ is the stellar density near our
Sun, $R_{l}=2.6{\rm kpc}$ \citep{Juric2008} is the galactic length
scale and $r_{\odot}=8{\rm kpc}$ is the distance of the Sun from
the galactic center. The mass of the perturber is taken to be $0.6M_{\odot}$
which is the average mass of a star in the galaxy. The velocity dispersion
is set to the velocity dispersion of the flat rotation curve of the
galaxy, namely $\sigma=50{\rm kms^{-1}}$.

\begin{figure}

\includegraphics[width=1.02\columnwidth]{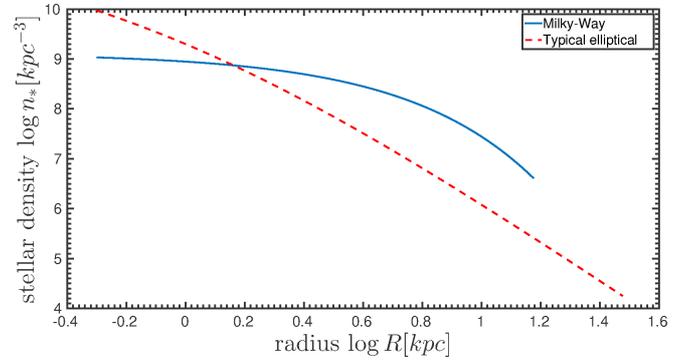}\caption{\label{fig:Numebr-stellar-density}Number stellar density as a function
of distance. The blue solid line represents the Milky-Way galaxy .
The red dashed line represents a typical elliptical galaxy.}

\end{figure}

For elliptical galaxies we take the density profile from \citep{Hernquist1990} and translate it to stellar density given an average stellar mass of $0.6M_{\odot}.$
\begin{equation}
\tilde{n}_{*}\left(r\right)=\frac{M_{{\rm galaxy}}}{2\pi r}\frac{r_{*}}{\left(r+r_{*}\right)^{3}}\label{eq:elliptical}
\end{equation}
where $r_{*}=1{\rm kpc}$ is the scale length of the galaxy, $M_{{\rm galaxy}}=10^{11}M_{\odot}$
is the total stellar (and not total) mass of the galaxy. The velocity
dispersion for a typical elliptical galaxy we consider is 
$\sigma=160{\rm kms^{-1}}.$
Figure \ref{fig:Numebr-stellar-density} shows the stellar density profiles
of the two prototypes of galaxies.

Next, we estimate the fraction of TBHs out of the entire stellar population,
$f_{{\rm TBH}}$. The fraction of BHs out of the entire stellar population
is $f_{{\rm primary}}\approx10^{-3}$ \citep{Kroupa2001}. We assume
all stars with masses greater than $20M_{\odot}$ turn into BHs without
any natal-kicks \citep[and others]{Belczynski2016,Mandel2016a} and
the triple fraction of all BHs progenitors is $f_{{\rm triple}}=0.25$ \citep{Sana2014}.
For the mass ratio distribution in the inner binary, $Q_{{\rm inner}}$ we consider a uniform distribution $Q_{{\rm inner}}\in\left(0.3,1\right)$ \citep{Moe2016},
this translates to a fraction of $f_{{\rm secondary}}\approx0.53$
of the inner binary companions to also form BHs. Additionally, we can compute
the fraction of the inner binaries (accounting for their total main-sequence
masses) to have a third BH progenitor companion, in this case we use
a different mass ratio distribution $Q_{{\rm outer}}\propto M_{{\rm binary}}^{-2}$
\citep{Moe2016}, as to apply for wide separation systems, and consider the range $Q_{{\rm outer}}\in\left(0.3,1\right)$. 
$M_{{\rm binary}}$ is the total mass of the inner binary. This translates
to the fraction of the tertiaries forming BHs of $f_{{\rm traitery}}\approx0.76$.
Hence the total fraction of TBHs from the stellar population is 
\begin{equation}
f_{{\rm TBH}}=f_{{\rm primary}}\times f_{{\rm secondary}}\times f_{{\rm tertiary}}\times f_{{\rm triple}}\approx1\times10^{-4}.
\end{equation}

The SMA distribution $f_{a_{1}}$and $f_{a_{2}}$is taken to be a
log-uniform, where the inner binary SMA ranges between $a_{1}\in\left(0.1{\rm AU,}100{\rm AU}\right)$
while $a_{2}\in\left(10^{3}{\rm AU},10^{5}{\rm AU}\right)$. Following
\citep{Michaely2019,Perets2012a,Igoshev2019} we take the overall fraction
of triple systems that are wider than $10^{3}{\rm AU}$ to be $f_{{\rm {\rm wide}}}=0.2$.

\subsection{Volumetric prompt-merger rates}
In the following we calculate the volumetric delayed-merger rate 
and the volumetric eccentric delayed-merger rate. 
\subsubsection*{Spiral galaxies}
For a MW-like galaxy the rate of BH prompt GW-mergers from perturbed wide TBH in the
field is just 
\begin{equation}
\Gamma_{{\rm MW}}=\int_{0.5{\rm kpc}}^{15{\rm kpc}}\int_{10^{3}{\rm AU}}^{10^{5}{\rm AU}}\int_{10^{-1}{\rm AU}}^{10^{2}{\rm AU}}\frac{L_{{\rm merger}}\left(a_{1},a_{2},n_{*}\right)}{10{\rm Gyr}}da_{1}da_{2}dN\left(r\right)\label{eq:MW_merger_IMS}
\end{equation}
\[
\approx0.017{\rm {\rm Myr^{-1}}}
\]
where $L_{{\rm merger}}\equiv L\left(a_{1},a_{2},n_{*}\right)f_{a_{1}}f_{a_{2}}f_{{\rm TBH}}f\left(a_{1}\right)_{{\rm merger}}$.
In order to translate this rate to the volumetric merger rate in spiral galaxies
$R_{{\rm spiral}}$, we follow \citep{Belczynski2016} and calculate the rate $R_{{\rm spiral}}=10^{3}n_{{\rm spiral}}\times\Gamma_{{\rm MW}}$, to get 
\begin{equation}
R_{{\rm spiral}}\approx0.2\times F_{{\rm model}}{\rm Gpc^{-3}yr^{-1}}.
\end{equation}
Where $n_{{\rm spiral}}=0.0116{\rm Mpc^{-3}}$ is the local density of
MW-like galaxies \citep{Kopparapu2008} and we define 
\begin{equation}
F_{{\rm model}}\equiv\left(\frac{f_{{\rm primary}}}{10^{-3}}\right)\left(\frac{f_{{\rm secondary}}}{0.53}\right)\left(\frac{f_{{\rm tritary}}}{0.76}\right)\left(\frac{f_{{\rm wide}}}{0.2}\right)\left(\frac{f_{{\rm triple}}}{0.25}\right)
\end{equation}
 to be the how the results depend on our model assumption. Moreover,
we can calculate the eccentric merger rate from this channel simply
by substituting $f_{{\rm merger}}$ with $f_{{\rm ecc}}$ from equation
(\ref{fig:ecc_distribution}), to find 
\begin{equation}
\Gamma_{{\rm MW,ecc}}=\int_{0.5{\rm kpc}}^{15{\rm kpc}}\int_{10^{3}{\rm AU}}^{10^{5}{\rm AU}}\int_{10^{-1}{\rm AU}}^{10^{2}{\rm AU}}\frac{L_{{\rm ecc}}}{10{\rm Gyr}}da_{1}da_{2}dN\left(r\right)\label{eq:MW_ec_IMS}
\end{equation}
\[
\approx0.01{\rm {\rm Myr^{-1}.}}
\]
where $L_{{\rm ecc}}\equiv L_{{\rm merger}}\left(a_{1},a_{2},n_{*}\right)f_{a_{1}}f_{a_{2}}f_{{\rm TBH}}f\left(a_{1}\right)_{{\rm eccntric}}$.
Hence, 
\begin{equation}
R_{{\rm spiral,ecc}}\approx0.12\times F_{{\rm model}}{\rm Gpc^{-3}yr^{-1}},\label{eq:R_ecc_IMS}
\end{equation}
and the fraction of eccentric mergers from this channel is consistent with $\sim78\%$.

\subsubsection*{Elliptical galaxies}

Following a similar procedure we now calculate the prompt-merger rate from elliptical galaxies.
Taking Eq. (\ref{eq:elliptical}) 
\begin{equation}
\Gamma_{{\rm elliptical}}=\int_{0.5{\rm kpc}}^{30{\rm kpc}}\int_{10^{3}{\rm AU}}^{10^{5}{\rm AU}}\int_{10^{-1}{\rm AU}}^{10^{2}{\rm AU}}\frac{L_{{\rm merger}}\left(a_{1},a_{2},\tilde{n}_{*}\right)}{10{\rm Gyr}}da_{1}da_{2}dN\left(r\right)
\end{equation}
\[
\approx0.03{\rm {\rm Myr^{-1}}.}
\]
Next we input the number density of elliptical galaxies in the local universe $n_{{\rm elliptical}}\approx0.1{\rm Mpc^{-3}}$\citep{Samsing2014}
and get

\begin{equation}
R_{{\rm elliptical}}=3.2\times \rm F_{{\rm model}}{\rm Gpc^{-3}yr^{-1}}.\label{eq:R_merger_elliptical_IMS}
\end{equation}
and for the eccentric mergers we get

\begin{equation}
R_{{\rm elliptical,ecc}}=1.2\times \rm F_{{\rm model}}{\rm Gpc^{-3}yr^{-1}}.\label{eq:R_eccentric_elliptical_IMS}
\end{equation}

Adding the contributions from both spiral and elliptical galaxies we get a total volumetric prompt-merger rate to be
\begin{equation}
R_{{\rm resonant}}=R_{{\rm spiral}}+R_{{\rm elliptical}}\approx3.4\times \rm F_{{\rm model}}{\rm Gpc^{-3}yr^{-1}}.
\end{equation}
and the volumetric eccentric prompt-mergers 
\begin{equation}
R_{{\rm resonant,ecc}}=R_{{\rm spiral,ecc}}+R_{{\rm elliptical,ecc}}\approx1.2\times \rm F_{{\rm model}}{\rm Gpc^{-3}yr^{-1}}\label{eq:Resosant_ecc}
\end{equation}

\subsection{Volumetric delayed merger rates}
In the following we calculate the volumetric delayed-merger rate 
and the volumetric eccentric delayed-merger rate. 
Following the same procedure described in
subsection \ref{subsec:Post-resonance-state} we calculate the merger
rate by substituting $f_{{\rm merger}}$ with $f_{{\rm delay}}$ from
equation (\ref{eq:f_delay}). As done in the previous section, we calculate
the merger rate of systems with $a_{1}\in\left(10^{-1}{\rm AU},10^{2}{\rm AU}\right)$
for both types of galaxies.

For spiral galaxies we find 
\begin{equation}
\Gamma_{{\rm spiral,delay}}\approx0.9{\rm {\rm Myr^{-1}}}
\end{equation}
which leads to 
\begin{equation}
R_{{\rm spiral,delay}}\approx9.2\times \rm F_{{\rm model}}{\rm Gpc^{-3}yr^{-1}}.
\end{equation}

For elliptical galaxies we find 
\begin{equation}
\Gamma_{{\rm elliptical,delay}}\approx1.5{\rm {\rm Myr^{-1}}}
\end{equation}
which translates to 
\begin{equation}
R_{{\rm elliptical,delay}}\approx150\times \rm F_{{\rm model}}{\rm Gpc^{-3}yr^{-1}}.
\end{equation}

In total we compute a volumetric rate of 
\begin{equation}
R_{{\rm delay}}=R_{{\rm spiral}}+R_{{\rm elliptical}}\approx160\times \rm F_{{\rm model}}{\rm Gpc^{-3}yr^{-1}}.
\end{equation}
Unlike the prompt-mergers only $\sim0.3\%$ of delayed-mergers end up as eccentric at $10{\rm Hz}$. Therefore we
expect only $\sim0.5\times F_{{\rm model}}{\rm Gpc^{-3}yr^{-1}}$
eccentric volumetric rate mergers from this channel.

\section{Discussion}
\label{sec:Discussion}
\subsection{Model assumptions}

The progenitor model for BBH GW-merger presented makes use and is  based on several assumptions,
in the following we address each them.

  \textbf{Natal kicks}. We cannot emphasize this point enough. This is the \textit{most} critical assumption we make is that the BHs discussed here, receive no natal kick at 
birth. The importance of this assumption, as we discussed in more depth in \citep{Michaely2019}, is that ultra-wide binaries/triples
are highly susceptible to disruption by such kicks. The binding energy of the outer
binary in the triple are very small, and natal-kicks of comparable velocity to the typical orbital velocity 
of the outer orbit or higher would disrupt the triple, significantly decreasing the number of potential 
TBH progenitors as discussed in \citep{Silsbee2017}. 

Currently, BH natal kicks are poorly
constrained \citep{Repetto2012,Repetto2017}. However, there is some evidence
that BH are formed following failed supernova (SN) \citep{Ertl2015,Adams2017}. In
the failed SN scenario large amount of fallback is accreted on the
newly formed compact object and suppresses any natal kicks, as the BH forms through direct collapse \citep{Fryer1999}. In fact, most if not all other theoretical models that can potentially reproduce the inferred high rates of
BBH GW-mergers follow similar assumptions, and also assume no or low-velocity natal-kicks
for BHs, or no/low natal kicks for higher mass BHs that form through direct collapse  \citep[e.g.][where many of the dynamical models make use of the same assumptions]{Belczynski2016,Belczynski2008,Belczynski2007}. We generally follow the same approach.

\textbf{Equal BH masses}. Here we considered only TBHs composed of same-mass BH components. 
This simplistic assumption is made as a
first step in developing this model. In the future we will expand
the mathematical formalism to account for unequal masses. Nevertheless,
we do not expect the rate to change dramatically \citep[e.g.][]{Samsing2018b}, when unequal BHs are considered. 

Nevertheless, we briefly discuss possible implications of our model in respect to the mass-function of the GW-mergers.
The masses of the component of the inner binaries are likely to be correlated, as we discussed in the assumptions regarding the rate calculations in section \ref{sec:GW-volume-rates}  and generally be more similar to those expected from the isolated binaries channel, where short period binaries serve as GW-sources progenitors. The outer third component, might be expected to be randomly drawn from a regular mass-function, as it forms almost in isolation, given the large separation from the inner binary (although in case it was dynamically captured, e.g. \citet{Perets2012a}, it would have some preference to higher masses).   
Since typically the less massive component is ejected in binary-single encounters, the dynamics will systematically give rise to overall higher mass BHs to take part in mergers compared with the BH mass function of single, or even isolated binary BHs. A more detailed prediction, however, will require further study of the binary-single encounter dynamics of unequal mass TBHs. 

\textbf{Inner binary SMA boundaries}. We set the lower boundary of
the inner SMA to be $a_{1}=0.1{\rm AU}.$ The reason is that the merger
time via GW emission of a circular binary with $m_{1}=m_{2}=10M_{\odot}$
at $a_{1}=0.1{\rm AU}$ is $t_{{\rm merger}}\approx10^{10}{\rm yr}$.
Hence, binaries with SMA smaller than $0.1{\rm AU}$ may merge in
isolation even without any perturbations, and thereby our model would not increase the GW-merger rates
originating from such short-period binaries. Nevertheless, as we discussed above, such binaries which
are part of TBHs such as those we discussed, might still evolve through the triple instability
we present here, and in this case their merger characteristics will differ, in particular 
their eccentricities might be higher, they will have shorter DTD, and should not generally show a spin-orbit alignment.

For completeness we present in table \ref{tab:merger_Rates_for_e-2}
the volumetric merger rate accounting for binaries with initial SMA
$a_{1}=10^{-2}{\rm AU}$ in order to compare with our results.

\begin{table*}

\caption{\label{tab:merger_Rates_for_e-2} The volumetric merger rates for
the case where $a_{1}\in\left(10^{-2}{\rm AU},10^{2}{\rm AU}\right)$.
Effectively these numbers include the isolated binary rates, hence
they do not represent and \emph{additional} contribution to the merger rates from the TBH channel.}

\begin{tabular}{|c|c|c|c|}
\hline 
 & prompt mergers $\rm F_{{\rm model}}{\rm Gpc^{-3}yr^{-1}}$ & eccentric mergers $\rm F_{{\rm model}}{\rm Gpc^{-3}yr^{-1}}$ & delayed mergers $\rm F_{{\rm model}}{\rm Gpc^{-3}yr^{-1}}$\tabularnewline
\hline 
\hline 
spirals & $\sim0.6$ & $\sim0.5$ & $\sim15$\tabularnewline
\hline 
elliptical & $\sim9.1$ & $\sim8.6$ & $\sim250$\tabularnewline
\hline 
\end{tabular}
\end{table*}

\textbf{Volumetric rate calculation}. In order to calculate the volumetric rate 
we make the assumption that the galaxy densities, both spiral and elliptical are $n_{\rm spiral}=$ and $n_{\rm elliptical}$. Furthermore, we assume the MW is the prototype of spiral
galaxies with velocity dispersion of $~50\rm kms^{-1}$ and total mass of $10^{11}M_\odot$  and the model we present in equation (\ref{eq:elliptical}) is the prototype for elliptical
with total mass of $10^{11}M_\odot$. This assumption may change the merger rate significantly for different galaxy prototypes. Specifically, for elliptical galaxies with total mass of 
$~5\times10^{10}M_{\odot}$ the rate decreases by an order of magnitude. The sensitivity of our results
by the specific model for a prototype galaxy motivates us to explore the issue in future research.
Moreover, in \citet{Michaely2019} we calculated the merger rate for wide binary systems solely for spiral galaxies. Following this work we 
argue that the rate of mergers from wide BBHs systems is governed from elliptical rather than spiral. We will explore this scenario elsewhere.

\subsubsection{Delay time distribution (DTD)}

\begin{figure}
\includegraphics[width=1.00\columnwidth]{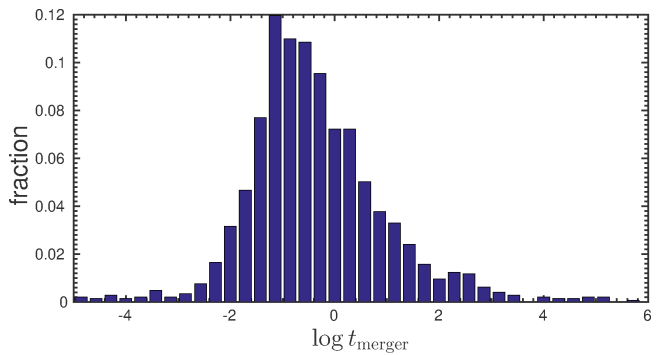}\caption{\label{fig:DTD}The distribution of merger times in the resonant phase.
The merger time, $t_{{\rm merger}}\lesssim P_{2}$ hence we regard
of mergers from the resonants stage as prompt once the TBH becomes
unstable}
\end{figure}

\citet{Michaely2019} showed that the DTD for wide binaries is uniform
in time. A priory one would expect that the DTD for the TBH case
might be more complicated due to the additional inspiral timescale during the 
resonant encounter or the later inspiral of the delayed mergers. 
In figure \ref{fig:DTD} we present the inspiral time for the prompt-mergers. 
We see that very short merger times
$t_{{\rm merger}}<10^{6}{\rm yr}$, which would hardly affect the overall uniform DTD
for the initial production of the destabilized TBHs, and can be generally neglected in that context.
The DTD for the prompt-GE channel is therefore expected to be generally uniform in time,
similar to the ultra-wide binary channel discussed in \citep{Michaely2019}.

The inspiral time of the delayed mergers is different  (see figure \ref{fig:f_merger}).
The distribution is dominated by a peak at $\sim10^{6}{\rm yr}$ approximately corresponding to the merger time of an initially circular binary with $a_{1}=0.01{\rm AU}$ which is the
most weighed value, given the assumed log uniform distribution of the SMAs. Considering a larger lower-bound $a_{1}>0.01{\rm AU}$ the peak would shift and be centered around $t\left(a_{1}\right)_{{\rm merger}}$, from equation (\ref{eq:t_merger})
until $a_{1}=0.1{\rm AU}$ which corresponds to a merger time of $\sim10^{10}{\rm yr}$
which is the upper cutoff. The shape to the right of the peak is effectively tracing the SMA distribution, $f_{a}$.

\begin{figure}
\includegraphics[width=1.0\columnwidth]{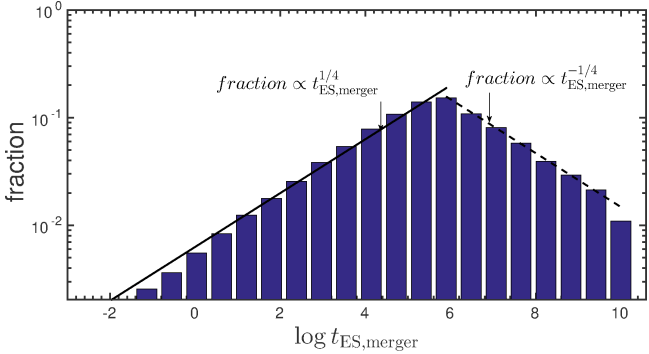}\caption{\label{fig:The-inspiral-time}
The inspiral time for all delayed mergers weighted by the distribution of the inner SMA, $a_{1}$.
Black solid line is to guide the eye with a slope that corresponds
to $t_{{\rm delay,merger}}^{1/4}$ and the black dashed line corresponds
to $t_{{\rm delay,merger}}^{-1/4}$}
\end{figure}

In this case the DTD would be slightly affected by the additional merger timescale for the shorter period binaries, 
and only somewhat change for the tail distribution of long merger times, giving rise to some modulation of the DTD, leading to a slightly decreasing DTD function. 

We should also briefly note that at the early stages of galaxy formation, in particular in disk galaxies, the initial stellar densities
and the number of BHs are initially low, compared to our basic model assumptions, accounting only for fully formed
galaxies. However, this should hardly affect the observable GW-sources most of which would not originate from such early times.

\subsubsection{Spin distribution}
The dynamical process of multiple binary-single interactions effectively 
samples the phase space chaotically such that the end state inclination distribution
is close to isotropic \citep{Stone2019b}. It is likely that this similarly 
holds for the resonant phase as well, where multiple though smaller number of 
encounters occur, although such assumption needs to be better verified in future studies. Overall we expect to find, similar
to dynamical mergers in  dense environments, an isotropic distribution 
of the orbital inclination and therefore an isotropic spin-orbit alignment distribution.
In that respect, the current findings of a preference for either an isotropic spin distribution or low spin magnitudes for the observed systems \citep{Will2014}, are consistent with our suggested channel. 

\subsubsection{Eccentric mergers distributions}
It was previously shown that eccentric mergers can originate from dynamical channels in dense cluster environments \citep[e.g.][]{Samsing2017,Rodriguez2018}, which predict
a volumetric rate of eccentric mergers of $\sim0.2-1{\rm Gpc^{-3}yr^{-1}}$.
As we show here, eccentric mergers can arise from the wide-TBH channel in the field. We find
in equation (\ref{eq:Resosant_ecc}) a volumetric rate of $\sim5\times \rm F_{{\rm model}}{\rm Gpc^{-3}yr^{-1}}$
eccentric mergers from the prompt-merger channel, which dominate the contribution of eccentric
mergers from this TBH scenario. Additionally, we present their distribution
in the inset of figure \ref{fig:ecc_distribution} and expect a peak
of the distribution with $\log\left(1-e_{1}\right)\approx-3.5$. This
extremely high eccentricities correspond to a unique signal in aLIGO/VIRGO.

Moreover, we find that, $0.3\%$ of the delayed-mergers we studied 
are eccentric at $10{\rm Hz}$ see figure \ref{fig:ecc_dist_endstate}.
This correspond to a rate of $\sim0.3-0.8\times \rm  F_{{\rm model}}{\rm Gpc^{-3}yr^{-1}}$.
Combining the prompt-merger and delayed-merger contributions we find an overall  rate of eccentric mergers of 
\begin{equation}
R_{{\rm eccentric,TBH}}\approx1-10\times \rm F_{{\rm model}}{\rm Gpc^{-3}yr^{-1}}.
\end{equation}

\section{Summary}
\label{sec:Summary}
In this paper we extended our previous study of BBH GW-sources formation from ultra-wide binaries perturbed by random flyby encounters in the field, and studied ultra-wide triples. We calculate the merger rate and eccentric merger rate of BBH originating from this channel and find them to potentially be one of the main channels for BBH GW sources. 

Wide TBH systems are gravitationally perturbed by random flybys of stars in the field, giving rise to a random walk of their outer-binary angular momentum. 
As a result a fraction
of the TBH outer-binaries become highly eccentric and their pericenter is sufficiently decreased to give rise to a strong encounter between the outer TBH component and the inner binary, destabilizing the system and driving it into an effective binary-single resonant encounter similar to such encounters that drive dynamical formation channels of GW-sources in dense stellar clusters. 
occurring in Consequently the TBHs evolve through a sequence of many binary-single encounters, during which two of the BHs might merger in what we term a prompt-GE mergers. Alternatively, one of the BHs could be ejected, leaving behind a remnant, more compact BBH. In the later case, sufficiently compact remnant BBHs would inspiral and merger through GW emission, and contribute to the formation of detectable GW sources, in what we term the delayed-mergers channel.  
We find the total volumetric rate of systems that merge via GW emission from both channels to be   
\begin{equation}
R_{{\rm IMS,merger}}\approx50-150{\rm \times F_{{\rm model}}Gpc^{-3}yr^{-1}}
\end{equation}
and an eccentric GW-mergers volumetric rate of 
\begin{equation}
R_{{\rm IMS,eccentric}}\approx1-10{\rm \times F_{{\rm model}}Gpc^{-3}yr^{-1}}.
\end{equation}
comparable and consistent with the currently inferred rate of BBH GW-mergers, and consistent with the current no detections of eccentric mergers, given the, still, too low statistics. We do expect, however, a few eccentric mergers to be detected over the coming few years, once the cumulative number of identified GW-sources is of the order of several hundreds. 

We also predict the spin-orbit alignment of the GW mergers from this channel to generally be isotropic. We also predict a close to uniform delay time distribution, with a significant contribution from both early and late type galaxies, and a preference for galaxies with higher velocity dispersions which are more favorable for field interactions with wide TBHs.

\section*{Acknowledgements}

EM thanks Johan Samsing for helpful and enlightening discussion that lead to this project.
HBP acknowledges support from the Kingsley distinguished-visitor program in Caltech, where some of the work has been done.  

\textbf{Data availability} The data underlying this article will be shared on reasonable request to the corresponding author.
\bibliographystyle{mnras}
\bibliography{TBH.bib}

\begin{thebibliography}{}
\makeatletter
\relax
\def\mn@urlcharsother{\let\do\@makeother \do\$\do\&\do\#\do\^\do\_\do\%\do\~}
\def\mn@doi{\begingroup\mn@urlcharsother \@ifnextchar [ {\mn@doi@}
  {\mn@doi@[]}}
\def\mn@doi@[#1]#2{\def\@tempa{#1}\ifx\@tempa\@empty \href
  {http://dx.doi.org/#2} {doi:#2}\else \href {http://dx.doi.org/#2} {#1}\fi
  \endgroup}
\def\mn@eprint#1#2{\mn@eprint@#1:#2::\@nil}
\def\mn@eprint@arXiv#1{\href {http://arxiv.org/abs/#1} {{\tt arXiv:#1}}}
\def\mn@eprint@dblp#1{\href {http://dblp.uni-trier.de/rec/bibtex/#1.xml}
  {dblp:#1}}
\def\mn@eprint@#1:#2:#3:#4\@nil{\def\@tempa {#1}\def\@tempb {#2}\def\@tempc
  {#3}\ifx \@tempc \@empty \let \@tempc \@tempb \let \@tempb \@tempa \fi \ifx
  \@tempb \@empty \def\@tempb {arXiv}\fi \@ifundefined
  {mn@eprint@\@tempb}{\@tempb:\@tempc}{\expandafter \expandafter \csname
  mn@eprint@\@tempb\endcsname \expandafter{\@tempc}}}

\bibitem[\protect\citeauthoryear{{Adams}, {Kochanek}, {Gerke}, {Stanek}  \&
  {Dai}}{{Adams} et~al.}{2017}]{Adams2017}
{Adams} S.~M.,  {Kochanek} C.~S.,  {Gerke} J.~R.,  {Stanek} K.~Z.,   {Dai} X.,
  2017, \mn@doi [\mnras] {10.1093/mnras/stx816}, \href
  {https://ui.adsabs.harvard.edu/abs/2017MNRAS.468.4968A} {468, 4968}

\bibitem[\protect\citeauthoryear{{Antognini}, {Shappee}, {Thompson}  \&
  {Amaro-Seoane}}{{Antognini} et~al.}{2014}]{Antognini2014}
{Antognini} J.~M.,  {Shappee} B.~J.,  {Thompson} T.~A.,   {Amaro-Seoane} P.,
  2014, \mn@doi [\mnras] {10.1093/mnras/stu039}, \href
  {http://adsabs.harvard.edu/abs/2014MNRAS.439.1079A} {439, 1079}

\bibitem[\protect\citeauthoryear{{Antonini} \& {Perets}}{{Antonini} \&
  {Perets}}{2012}]{Antonini2012}
{Antonini} F.,  {Perets} H.~B.,  2012, \mn@doi [\apj]
  {10.1088/0004-637X/757/1/27}, \href
  {http://adsabs.harvard.edu/abs/2012ApJ...757...27A} {757, 27}

\bibitem[\protect\citeauthoryear{{Antonini}, {Murray}  \& {Mikkola}}{{Antonini}
  et~al.}{2014}]{Antonini2014}
{Antonini} F.,  {Murray} N.,   {Mikkola} S.,  2014, \mn@doi [\apj]
  {10.1088/0004-637X/781/1/45}, \href
  {http://adsabs.harvard.edu/abs/2014ApJ...781...45A} {781, 45}

\bibitem[\protect\citeauthoryear{{Antonini}, {Chatterjee}, {Rodriguez},
  {Morscher}, {Pattabiraman}, {Kalogera}  \& {Rasio}}{{Antonini}
  et~al.}{2016}]{Antonini2016}
{Antonini} F.,  {Chatterjee} S.,  {Rodriguez} C.~L.,  {Morscher} M.,
  {Pattabiraman} B.,  {Kalogera} V.,   {Rasio} F.~A.,  2016, \mn@doi [\apj]
  {10.3847/0004-637X/816/2/65}, \href
  {https://ui.adsabs.harvard.edu/abs/2016ApJ...816...65A} {816, 65}

\bibitem[\protect\citeauthoryear{{Antonini}, {Toonen}  \& {Hamers}}{{Antonini}
  et~al.}{2017}]{Antonini2017}
{Antonini} F.,  {Toonen} S.,   {Hamers} A.~S.,  2017, \mn@doi [\apj]
  {10.3847/1538-4357/aa6f5e}, \href
  {http://adsabs.harvard.edu/abs/2017ApJ...841...77A} {841, 77}

\bibitem[\protect\citeauthoryear{{Bahcall}, {Hut}  \& {Tremaine}}{{Bahcall}
  et~al.}{1985}]{Bahcall1985}
{Bahcall} J.~N.,  {Hut} P.,   {Tremaine} S.,  1985, \mn@doi [\apj]
  {10.1086/162953}, \href {http://adsabs.harvard.edu/abs/1985ApJ...290...15B}
  {290, 15}

\bibitem[\protect\citeauthoryear{{Banerjee}}{{Banerjee}}{2018}]{Banerjee2018}
{Banerjee} S.,  2018, \mn@doi [\mnras] {10.1093/mnras/stx2347}, \href
  {https://ui.adsabs.harvard.edu/abs/2018MNRAS.473..909B} {473, 909}

\bibitem[\protect\citeauthoryear{{Belczynski}, {Kalogera}  \&
  {Bulik}}{{Belczynski} et~al.}{2002}]{Belczynski2002}
{Belczynski} K.,  {Kalogera} V.,   {Bulik} T.,  2002, \mn@doi [\apj]
  {10.1086/340304}, \href {http://adsabs.harvard.edu/abs/2002ApJ...572..407B}
  {572, 407}

\bibitem[\protect\citeauthoryear{{Belczynski}, {Sadowski}  \&
  {Rasio}}{{Belczynski} et~al.}{2004}]{Belczynski2004}
{Belczynski} K.,  {Sadowski} A.,   {Rasio} F.~A.,  2004, \mn@doi [\apj]
  {10.1086/422191}, \href {http://adsabs.harvard.edu/abs/2004ApJ...611.1068B}
  {611, 1068}

\bibitem[\protect\citeauthoryear{{Belczynski}, {Taam}, {Kalogera}, {Rasio}  \&
  {Bulik}}{{Belczynski} et~al.}{2007}]{Belczynski2007}
{Belczynski} K.,  {Taam} R.~E.,  {Kalogera} V.,  {Rasio} F.~A.,   {Bulik} T.,
  2007, \mn@doi [\apj] {10.1086/513562}, \href
  {http://adsabs.harvard.edu/abs/2007ApJ...662..504B} {662, 504}

\bibitem[\protect\citeauthoryear{{Belczynski}, {Kalogera}, {Rasio}, {Taam},
  {Zezas}, {Bulik}, {Maccarone}  \& {Ivanova}}{{Belczynski}
  et~al.}{2008}]{Belczynski2008}
{Belczynski} K.,  {Kalogera} V.,  {Rasio} F.~A.,  {Taam} R.~E.,  {Zezas} A.,
  {Bulik} T.,  {Maccarone} T.~J.,   {Ivanova} N.,  2008, \mn@doi [\apjs]
  {10.1086/521026}, \href {http://adsabs.harvard.edu/abs/2008ApJS..174..223B}
  {174, 223}

\bibitem[\protect\citeauthoryear{{Belczynski}, {Repetto}, {Holz},
  {O'Shaughnessy}, {Bulik}, {Berti}, {Fryer}  \& {Dominik}}{{Belczynski}
  et~al.}{2016}]{Belczynski2016}
{Belczynski} K.,  {Repetto} S.,  {Holz} D.~E.,  {O'Shaughnessy} R.,  {Bulik}
  T.,  {Berti} E.,  {Fryer} C.,   {Dominik} M.,  2016, \mn@doi [\apj]
  {10.3847/0004-637X/819/2/108}, \href
  {http://adsabs.harvard.edu/abs/2016ApJ...819..108B} {819, 108}

\bibitem[\protect\citeauthoryear{{Dominik}, {Belczynski}, {Fryer}, {Holz},
  {Berti}, {Bulik}, {Mandel}  \& {O'Shaughnessy}}{{Dominik}
  et~al.}{2012}]{Dominik2012}
{Dominik} M.,  {Belczynski} K.,  {Fryer} C.,  {Holz} D.~E.,  {Berti} E.,
  {Bulik} T.,  {Mandel} I.,   {O'Shaughnessy} R.,  2012, \mn@doi [\apj]
  {10.1088/0004-637X/759/1/52}, \href
  {http://adsabs.harvard.edu/abs/2012ApJ...759...52D} {759, 52}

\bibitem[\protect\citeauthoryear{{Dominik} et~al.,}{{Dominik}
  et~al.}{2015}]{Dominik2015}
{Dominik} M.,  et~al., 2015, \mn@doi [\apj] {10.1088/0004-637X/806/2/263},
  \href {http://adsabs.harvard.edu/abs/2015ApJ...806..263D} {806, 263}

\bibitem[\protect\citeauthoryear{{Ertl}, {Janka}, {Woosley}, {Sukhbold}  \&
  {Ugliano}}{{Ertl} et~al.}{2015}]{Ertl2015}
{Ertl} T.,  {Janka} H.-T.,  {Woosley} S.~E.,  {Sukhbold} T.,   {Ugliano} M.,
  2015, preprint, \href {http://adsabs.harvard.edu/abs/2015arXiv150307522E} {}
  (\mn@eprint {arXiv} {1503.07522})

\bibitem[\protect\citeauthoryear{{Fragione} \& {Kocsis}}{{Fragione} \&
  {Kocsis}}{2018}]{Fragione2018}
{Fragione} G.,  {Kocsis} B.,  2018, \mn@doi [\prl]
  {10.1103/PhysRevLett.121.161103}, \href
  {https://ui.adsabs.harvard.edu/abs/2018PhRvL.121p1103F} {121, 161103}

\bibitem[\protect\citeauthoryear{{Fragione}, {Grishin}, {Leigh}, {Perets}  \&
  {Perna}}{{Fragione} et~al.}{2019}]{Fragione2019}
{Fragione} G.,  {Grishin} E.,  {Leigh} N. W.~C.,  {Perets} H.~B.,   {Perna} R.,
   2019, \mn@doi [\mnras] {10.1093/mnras/stz1651}, \href
  {https://ui.adsabs.harvard.edu/abs/2019MNRAS.488...47F} {488, 47}

\bibitem[\protect\citeauthoryear{{Fryer}, {Woosley}  \& {Hartmann}}{{Fryer}
  et~al.}{1999}]{Fryer1999}
{Fryer} C.~L.,  {Woosley} S.~E.,   {Hartmann} D.~H.,  1999, \mn@doi [\apj]
  {10.1086/307992}, \href {http://adsabs.harvard.edu/abs/1999ApJ...526..152F}
  {526, 152}

\bibitem[\protect\citeauthoryear{{Hamers}, {Bar-Or}, {Petrovich}  \&
  {Antonini}}{{Hamers} et~al.}{2018}]{Hamers2018}
{Hamers} A.~S.,  {Bar-Or} B.,  {Petrovich} C.,   {Antonini} F.,  2018, \mn@doi
  [\apj] {10.3847/1538-4357/aadae2}, \href
  {https://ui.adsabs.harvard.edu/abs/2018ApJ...865....2H} {865, 2}

\bibitem[\protect\citeauthoryear{{Hamilton} \& {Rafikov}}{{Hamilton} \&
  {Rafikov}}{2019}]{Hamilton2019}
{Hamilton} C.,  {Rafikov} R.~R.,  2019, \mn@doi [\mnras]
  {10.1093/mnras/stz2026}, \href
  {https://ui.adsabs.harvard.edu/abs/2019MNRAS.488.5512H} {488, 5512}

\bibitem[\protect\citeauthoryear{{Harry} \& {LIGO Scientific
  Collaboration}}{{Harry} \& {LIGO Scientific Collaboration}}{2010}]{Harry2010}
{Harry} G.~M.,  {LIGO Scientific Collaboration} 2010, \mn@doi [Classical and
  Quantum Gravity] {10.1088/0264-9381/27/8/084006}, \href
  {https://ui.adsabs.harvard.edu/abs/2010CQGra..27h4006H} {27, 084006}

\bibitem[\protect\citeauthoryear{{Heggie}}{{Heggie}}{1975}]{Heggie1975}
{Heggie} D.~C.,  1975, \mn@doi [\mnras] {10.1093/mnras/173.3.729}, \href
  {https://ui.adsabs.harvard.edu/abs/1975MNRAS.173..729H} {173, 729}

\bibitem[\protect\citeauthoryear{{Hernquist}}{{Hernquist}}{1990}]{Hernquist1990}
{Hernquist} L.,  1990, \mn@doi [\apj] {10.1086/168845}, \href
  {https://ui.adsabs.harvard.edu/abs/1990ApJ...356..359H} {356, 359}

\bibitem[\protect\citeauthoryear{{Hills}}{{Hills}}{1975}]{Hills1975}
{Hills} J.~G.,  1975, \mn@doi [\aj] {10.1086/111815}, \href
  {https://ui.adsabs.harvard.edu/abs/1975AJ.....80..809H} {80, 809}

\bibitem[\protect\citeauthoryear{{Hills}}{{Hills}}{1981}]{Hills1981}
{Hills} J.~G.,  1981, \mn@doi [\aj] {10.1086/113058}, \href
  {http://adsabs.harvard.edu/abs/1981AJ.....86.1730H} {86, 1730}

\bibitem[\protect\citeauthoryear{{Hoang}, {Naoz}, {Kocsis}, {Rasio}  \&
  {Dosopoulou}}{{Hoang} et~al.}{2018}]{Hoang2018}
{Hoang} B.-M.,  {Naoz} S.,  {Kocsis} B.,  {Rasio} F.~A.,   {Dosopoulou} F.,
  2018, \mn@doi [\apj] {10.3847/1538-4357/aaafce}, \href
  {https://ui.adsabs.harvard.edu/abs/2018ApJ...856..140H} {856, 140}

\bibitem[\protect\citeauthoryear{{Hut} \& {Tremaine}}{{Hut} \&
  {Tremaine}}{1985}]{Hut1985}
{Hut} P.,  {Tremaine} S.,  1985, \mn@doi [\aj] {10.1086/113868}, \href
  {https://ui.adsabs.harvard.edu/abs/1985AJ.....90.1548H} {90, 1548}

\bibitem[\protect\citeauthoryear{{Igoshev} \& {Perets}}{{Igoshev} \&
  {Perets}}{2019}]{Igoshev2019}
{Igoshev} A.~P.,  {Perets} H.~B.,  2019, arXiv e-prints, \href
  {http://adsabs.harvard.edu/abs/2019arXiv190105972I} {}

\bibitem[\protect\citeauthoryear{{Juri{\'c}} et~al.,}{{Juri{\'c}}
  et~al.}{2008}]{Juric2008}
{Juri{\'c}} M.,  et~al., 2008, \mn@doi [\apj] {10.1086/523619}, \href
  {http://adsabs.harvard.edu/abs/2008ApJ...673..864J} {673, 864}

\bibitem[\protect\citeauthoryear{{Kaib} \& {Raymond}}{{Kaib} \&
  {Raymond}}{2014}]{Kaib2014}
{Kaib} N.~A.,  {Raymond} S.~N.,  2014, \mn@doi [\apj]
  {10.1088/0004-637X/782/2/60}, \href
  {http://adsabs.harvard.edu/abs/2014ApJ...782...60K} {782, 60}

\bibitem[\protect\citeauthoryear{{Kopparapu}, {Hanna}, {Kalogera},
  {O'Shaughnessy}, {Gonz{\'a}lez}, {Brady}  \& {Fairhurst}}{{Kopparapu}
  et~al.}{2008}]{Kopparapu2008}
{Kopparapu} R.~K.,  {Hanna} C.,  {Kalogera} V.,  {O'Shaughnessy} R.,
  {Gonz{\'a}lez} G.,  {Brady} P.~R.,   {Fairhurst} S.,  2008, \mn@doi [\apj]
  {10.1086/527348}, \href {http://adsabs.harvard.edu/abs/2008ApJ...675.1459K}
  {675, 1459}

\bibitem[\protect\citeauthoryear{{Kozai}}{{Kozai}}{1962}]{koz62}
{Kozai} Y.,  1962, \mn@doi [\aj] {10.1086/108790}, \href
  {http://adsabs.harvard.edu/abs/1962AJ.....67..591K} {67, 591}

\bibitem[\protect\citeauthoryear{{Kroupa}}{{Kroupa}}{2001}]{Kroupa2001}
{Kroupa} P.,  2001, \mn@doi [\mnras] {10.1046/j.1365-8711.2001.04022.x}, \href
  {http://adsabs.harvard.edu/abs/2001MNRAS.322..231K} {322, 231}

\bibitem[\protect\citeauthoryear{{LIGO Scientific Collaboration} \& {Virgo
  Collaboration}}{{LIGO Scientific Collaboration} \& {Virgo
  Collaboration}}{2019}]{Abbott2019}
{LIGO Scientific Collaboration} {Virgo Collaboration} 2019, \mn@doi [Physical
  Review X] {10.1103/PhysRevX.9.031040}, \href
  {https://ui.adsabs.harvard.edu/abs/2019PhRvX...9c1040A} {9, 031040}

\bibitem[\protect\citeauthoryear{{Leigh} et~al.,}{{Leigh}
  et~al.}{2018}]{Leigh2018}
{Leigh} N.~W.~C.,  et~al., 2018, \mn@doi [\mnras] {10.1093/mnras/stx3134},
  \href {http://adsabs.harvard.edu/abs/2018MNRAS.474.5672L} {474, 5672}

\bibitem[\protect\citeauthoryear{{Lidov}}{{Lidov}}{1962}]{Lidov1962}
{Lidov} M.~L.,  1962, \mn@doi [\planss] {10.1016/0032-0633(62)90129-0}, \href
  {http://adsabs.harvard.edu/abs/1962P%26SS....9..719L} {9, 719}

\bibitem[\protect\citeauthoryear{{Lightman} \& {Shapiro}}{{Lightman} \&
  {Shapiro}}{1977}]{Lightman1977}
{Lightman} A.~P.,  {Shapiro} S.~L.,  1977, \mn@doi [\apj] {10.1086/154925},
  \href {https://ui.adsabs.harvard.edu/abs/1977ApJ...211..244L} {211, 244}

\bibitem[\protect\citeauthoryear{{Mandel}}{{Mandel}}{2016}]{Mandel2016a}
{Mandel} I.,  2016, \mn@doi [\mnras] {10.1093/mnras/stv2733}, \href
  {http://adsabs.harvard.edu/abs/2016MNRAS.456..578M} {456, 578}

\bibitem[\protect\citeauthoryear{{Mandel} \& {de Mink}}{{Mandel} \& {de
  Mink}}{2016}]{Mandel2016}
{Mandel} I.,  {de Mink} S.~E.,  2016, \mn@doi [\mnras] {10.1093/mnras/stw379},
  \href {http://adsabs.harvard.edu/abs/2016MNRAS.458.2634M} {458, 2634}

\bibitem[\protect\citeauthoryear{{Mardling} \& {Aarseth}}{{Mardling} \&
  {Aarseth}}{2001}]{Mardling2001}
{Mardling} R.~A.,  {Aarseth} S.~J.,  2001, \mn@doi [\mnras]
  {10.1046/j.1365-8711.2001.03974.x}, \href
  {http://adsabs.harvard.edu/abs/2001MNRAS.321..398M} {321, 398}

\bibitem[\protect\citeauthoryear{{Merritt}}{{Merritt}}{2013}]{Merritt2013}
{Merritt} D.,  2013, \mn@doi [Classical and Quantum Gravity]
  {10.1088/0264-9381/30/24/244005}, \href
  {https://ui.adsabs.harvard.edu/abs/2013CQGra..30x4005M} {30, 244005}

\bibitem[\protect\citeauthoryear{{Michaely} \& {Perets}}{{Michaely} \&
  {Perets}}{2014}]{Michaely2014}
{Michaely} E.,  {Perets} H.~B.,  2014, \mn@doi [\apj]
  {10.1088/0004-637X/794/2/122}, \href
  {http://cdsads.u-strasbg.fr/abs/2014ApJ...794..122M} {794, 122}

\bibitem[\protect\citeauthoryear{{Michaely} \& {Perets}}{{Michaely} \&
  {Perets}}{2016}]{Michaely2016}
{Michaely} E.,  {Perets} H.~B.,  2016, \mn@doi [\mnras] {10.1093/mnras/stw368},
  \href {http://adsabs.harvard.edu/abs/2016MNRAS.458.4188M} {458, 4188}

\bibitem[\protect\citeauthoryear{{Michaely} \& {Perets}}{{Michaely} \&
  {Perets}}{2018}]{Michaely2018}
{Michaely} E.,  {Perets} H.~B.,  2018, \mn@doi [\apjl]
  {10.3847/2041-8213/aaacfc}, \href
  {https://ui.adsabs.harvard.edu/abs/2018ApJ...855L..12M} {855, L12}

\bibitem[\protect\citeauthoryear{{Michaely} \& {Perets}}{{Michaely} \&
  {Perets}}{2019}]{Michaely2019}
{Michaely} E.,  {Perets} H.~B.,  2019, \mn@doi [\apjl]
  {10.3847/2041-8213/ab5b9b}, \href
  {https://ui.adsabs.harvard.edu/abs/2019ApJ...887L..36M} {887, L36}

\bibitem[\protect\citeauthoryear{{Moe} \& {Di Stefano}}{{Moe} \& {Di
  Stefano}}{2016}]{Moe2016}
{Moe} M.,  {Di Stefano} R.,  2016, preprint, \href
  {http://adsabs.harvard.edu/abs/2016arXiv160605347M} {} (\mn@eprint {arXiv}
  {1606.05347})

\bibitem[\protect\citeauthoryear{{Naoz}}{{Naoz}}{2016}]{Naoz2016}
{Naoz} S.,  2016, \mn@doi [\araa] {10.1146/annurev-astro-081915-023315}, \href
  {https://ui.adsabs.harvard.edu/abs/2016ARA&A..54..441N} {54, 441}

\bibitem[\protect\citeauthoryear{{Perets} \& {Kouwenhoven}}{{Perets} \&
  {Kouwenhoven}}{2012}]{Perets2012a}
{Perets} H.~B.,  {Kouwenhoven} M.~B.~N.,  2012, \mn@doi [\apj]
  {10.1088/0004-637X/750/1/83}, \href
  {http://adsabs.harvard.edu/abs/2012ApJ...750...83P} {750, 83}

\bibitem[\protect\citeauthoryear{{Peters}}{{Peters}}{1964}]{Pet64}
{Peters} P.~C.,  1964, \mn@doi [Physical Review] {10.1103/PhysRev.136.B1224},
  \href {http://adsabs.harvard.edu/abs/1964PhRv..136.1224P} {136, 1224}

\bibitem[\protect\citeauthoryear{{Petrovich} \& {Antonini}}{{Petrovich} \&
  {Antonini}}{2017}]{Petrovich2017}
{Petrovich} C.,  {Antonini} F.,  2017, \mn@doi [\apj]
  {10.3847/1538-4357/aa8628}, \href
  {http://adsabs.harvard.edu/abs/2017ApJ...846..146P} {846, 146}

\bibitem[\protect\citeauthoryear{{Repetto}, {Davies}  \&
  {Sigurdsson}}{{Repetto} et~al.}{2012}]{Repetto2012}
{Repetto} S.,  {Davies} M.~B.,   {Sigurdsson} S.,  2012, \mn@doi [\mnras]
  {10.1111/j.1365-2966.2012.21549.x}, \href
  {http://adsabs.harvard.edu/abs/2012MNRAS.425.2799R} {425, 2799}

\bibitem[\protect\citeauthoryear{{Repetto}, {Igoshev}  \& {Nelemans}}{{Repetto}
  et~al.}{2017}]{Repetto2017}
{Repetto} S.,  {Igoshev} A.~P.,   {Nelemans} G.,  2017, \mn@doi [\mnras]
  {10.1093/mnras/stx027}, \href
  {http://adsabs.harvard.edu/abs/2017MNRAS.467..298R} {467, 298}

\bibitem[\protect\citeauthoryear{{Rodriguez}, {Chatterjee}  \&
  {Rasio}}{{Rodriguez} et~al.}{2016}]{Rodriguez2016}
{Rodriguez} C.~L.,  {Chatterjee} S.,   {Rasio} F.~A.,  2016, \mn@doi [\prd]
  {10.1103/PhysRevD.93.084029}, \href
  {http://adsabs.harvard.edu/abs/2016PhRvD..93h4029R} {93, 084029}

\bibitem[\protect\citeauthoryear{{Rodriguez}, {Amaro-Seoane}, {Chatterjee},
  {Kremer}, {Rasio}, {Samsing}, {Ye}  \& {Zevin}}{{Rodriguez}
  et~al.}{2018}]{Rodriguez2018}
{Rodriguez} C.~L.,  {Amaro-Seoane} P.,  {Chatterjee} S.,  {Kremer} K.,  {Rasio}
  F.~A.,  {Samsing} J.,  {Ye} C.~S.,   {Zevin} M.,  2018, \mn@doi [\prd]
  {10.1103/PhysRevD.98.123005}, \href
  {http://adsabs.harvard.edu/abs/2018PhRvD..98l3005R} {98, 123005}

\bibitem[\protect\citeauthoryear{{Samsing} \& {D'Orazio}}{{Samsing} \&
  {D'Orazio}}{2018}]{Samsing2018}
{Samsing} J.,  {D'Orazio} D.~J.,  2018, \mn@doi [\mnras]
  {10.1093/mnras/sty2334}, \href
  {http://adsabs.harvard.edu/abs/2018MNRAS.481.5445S} {481, 5445}

\bibitem[\protect\citeauthoryear{{Samsing} \& {Ramirez-Ruiz}}{{Samsing} \&
  {Ramirez-Ruiz}}{2017}]{Samsing2017}
{Samsing} J.,  {Ramirez-Ruiz} E.,  2017, \mn@doi [\apjl]
  {10.3847/2041-8213/aa6f0b}, \href
  {https://ui.adsabs.harvard.edu/abs/2017ApJ...840L..14S} {840, L14}

\bibitem[\protect\citeauthoryear{{Samsing}, {MacLeod}  \&
  {Ramirez-Ruiz}}{{Samsing} et~al.}{2014}]{Samsing2014}
{Samsing} J.,  {MacLeod} M.,   {Ramirez-Ruiz} E.,  2014, \mn@doi [\apj]
  {10.1088/0004-637X/784/1/71}, \href
  {https://ui.adsabs.harvard.edu/abs/2014ApJ...784...71S} {784, 71}

\bibitem[\protect\citeauthoryear{{Samsing}, {MacLeod}  \&
  {Ramirez-Ruiz}}{{Samsing} et~al.}{2018a}]{Samsing2018b}
{Samsing} J.,  {MacLeod} M.,   {Ramirez-Ruiz} E.,  2018a, \mn@doi [\apj]
  {10.3847/1538-4357/aaa715}, \href
  {https://ui.adsabs.harvard.edu/abs/2018ApJ...853..140S} {853, 140}

\bibitem[\protect\citeauthoryear{{Samsing}, {Askar}  \& {Giersz}}{{Samsing}
  et~al.}{2018b}]{Samsing2018c}
{Samsing} J.,  {Askar} A.,   {Giersz} M.,  2018b, \mn@doi [\apj]
  {10.3847/1538-4357/aaab52}, \href
  {https://ui.adsabs.harvard.edu/abs/2018ApJ...855..124S} {855, 124}

\bibitem[\protect\citeauthoryear{{Sana} et~al.,}{{Sana}
  et~al.}{2014}]{Sana2014}
{Sana} H.,  et~al., 2014, \mn@doi [\apjs] {10.1088/0067-0049/215/1/15}, \href
  {http://adsabs.harvard.edu/abs/2014ApJS..215...15S} {215, 15}

\bibitem[\protect\citeauthoryear{{Silsbee} \& {Tremaine}}{{Silsbee} \&
  {Tremaine}}{2017}]{Silsbee2017}
{Silsbee} K.,  {Tremaine} S.,  2017, \mn@doi [\apj] {10.3847/1538-4357/aa5729},
  \href {https://ui.adsabs.harvard.edu/abs/2017ApJ...836...39S} {836, 39}

\bibitem[\protect\citeauthoryear{{Stone} \& {Leigh}}{{Stone} \&
  {Leigh}}{2019}]{Stone2019b}
{Stone} N.~C.,  {Leigh} N. W.~C.,  2019, \mn@doi [\nat]
  {10.1038/s41586-019-1833-8}, \href
  {https://ui.adsabs.harvard.edu/abs/2019Natur.576..406S} {576, 406}

\bibitem[\protect\citeauthoryear{{Wen}}{{Wen}}{2003}]{Wen2003}
{Wen} L.,  2003, \mn@doi [\apj] {10.1086/378794}, \href
  {https://ui.adsabs.harvard.edu/abs/2003ApJ...598..419W} {598, 419}

\bibitem[\protect\citeauthoryear{{Will}}{{Will}}{2014}]{Will2014}
{Will} C.~M.,  2014, \mn@doi [\prd] {10.1103/PhysRevD.89.044043}, \href
  {http://adsabs.harvard.edu/abs/2014PhRvD..89d4043W} {89, 044043}

\bibitem[\protect\citeauthoryear{{de Mink} \& {Belczynski}}{{de Mink} \&
  {Belczynski}}{2015}]{deMink2015}
{de Mink} S.~E.,  {Belczynski} K.,  2015, \mn@doi [\apj]
  {10.1088/0004-637X/814/1/58}, \href
  {http://adsabs.harvard.edu/abs/2015ApJ...814...58D} {814, 58}

\makeatother
\end{thebibliography}

\end{document}